# Isoscalar and isovector giant resonances in $^{44}$Ca, $^{54}$Fe, $^{64,68}$Zn and $^{56,58,60,68}$Ni

G. Bonasera[1,*], S. Shlomo[1,+], D.H. Youngblood[1], Y.-W. Lui[1], J. Button[1], and X. Chen[2]

[1]Cyclotron Institute, Texas A&M University, College Station, Texas 77843, USA

[2]Department of Radiation Oncology, Medical College of Wisconsin, Milwaukee, Wisconsin 53226, USA

(Dated: July 2020)

We have studied the uncharacteristic behavior of the measured values of the isoscalar and isovector centroid energies, $E_{CEN}$, of giant resonances with multipolarity L=0-3 in $^{44}$Ca, $^{54}$Fe, $^{64,68}$Zn and $^{56,58,60,68}$Ni. For this purpose, we carried out calculations of $E_{CEN}$ within the spherical Hartree-Fock (HF)-based random phase approximation (RPA) theory with 33 different Skyrme-type effective nucleon-nucleon interactions. We have also determined the Pearson linear correlation coefficients between the calculated centroid energies and the various nuclear matter (NM) properties associated with each interaction and determined the sensitivity of $E_{CEN}$ to NM properties. We compared the calculated centroid energies of the giant resonances with experimental data and discuss the results. We note in particular, that we obtain good agreement between the calculated $E_{CEN}$ of isovector giant dipole resonance and the available experimental data.

---

[*] giacomo90@email.tamu.edu

[+] s-shlomo@tamu.edu

# I. INTRODUCTION

Nuclear giant resonances are an example of collective motion in the atomic nucleus and have been studied for many decades [1–5]. Nuclear giant resonances are classified in two modes of oscillation, neutrons and protons moving in-phase with each other (T=0, isoscalar) or out-of-phase (T=1, isovector) and of various multipolarities (L=0 monopole, L=1 dipole, and so on). The goal of these studies has been to determine, with ever increasing accuracy, the values of bulk nuclear matter (NM) properties in order to constrain the adopted form of the energy density functional (EDF) as well as the nuclear force. The improved EDFs can then be used to determine better equations of state of NM and to calculate properties of nuclei at and away from stability, properties of nuclear structure, the evolution of astrophysical objects and of heavy-ion collisions [6–10].

Similar to our previous studies, in which we analyzed the behavior of giant resonances across a wide range of masses from $^{40}$Ca to $^{208}$Pb [11] and then focused on the region A = 90-100 for isotopes of Mo and Zr [12–14], we now focus on the lower-mass region containing the $^{44}$Ca, $^{54}$Fe, $^{64,68}$Zn and $^{56,58,60,68}$Ni nuclei. We carried out calculations of centroid energies, $E_{CEN}$, within the Hartree-Fock (HF)-based random phase approximation (RPA) using 33 Skyrme-type effective interactions [15–17], for both isoscalar and isovector giant resonances of multipolarities of L = 0–3. We also compare our results with available experimental data including that recently obtained at Texas A&M University for $^{44}$Ca, $^{54}$Fe, $^{64,68}$Zn [18,19] as well as older data for the $^{58,60}$Ni isotopes [20]. This investigation is an extension of our study [18,19] involving these nuclei in which we presented experimental data as well as calculations employing only the KDE0v1 [21] Skyrme-type interaction and compared both strength functions and centroid energies to the experimental data.

In the following section we briefly describe the theoretical background for determining the centroid energies, $E_{CEN}$, of isoscalar and isovector giant resonances of multipolarity L = 0-3. Our results are compared with experimental data in section III where we present the calculated values of $E_{CEN}$ for all nuclei for each giant resonance separately. Here we also investigate the sensitivity of the values of $E_{CEN}$ to NM properties by calculating the corresponding Pearson linear correlation coefficients. Summary and conclusions are given in section IV.

## II. METHOD

In our calculations we adopt the standard (10-parameter) Skyrme effective nucleon-nucleon interaction [22]:

$$V_{ij} = t_0(1 + x_0 P_{ij}^\sigma)\delta(\vec{r}_i - \vec{r}_j) + \frac{1}{2}t_1(1 + x_1 P_{ij}^\sigma)[\overleftarrow{k}_{ij}^2 \delta(\vec{r}_i - \vec{r}_j) + \delta(\vec{r}_i - \vec{r}_j)\vec{k}_{ij}^2]$$
$$+ t_2(1 + x_2 P_{ij}^\sigma)\overleftarrow{k}_{ij}\delta(\vec{r}_i - \vec{r}_j)\vec{k}_{ij} + \frac{1}{6}t_3(1 + x_3 P_{ij}^\sigma)\rho^\alpha\left(\frac{\vec{r}_i + \vec{r}_j}{2}\right)\delta(\vec{r}_i - \vec{r}_j) \quad (1)$$
$$+ iW_0 \overleftarrow{k}_{ij}\delta(\vec{r}_i - \vec{r}_j)(\vec{\sigma}_1 + \vec{\sigma}_2) \times \vec{k}_{ij},$$

where $t_i$, $x_i$, $W_0$ and $\alpha$ are the Skyrme parameters listed in TABLE I of reference [11]. The spin exchange operator is given by $P_{ij}^\sigma$, the Pauli spin operator is $\vec{\sigma}_i$ and lastly the left and right momentum operators are given by $\vec{k}_{ij} = -\frac{i(\vec{\nabla}_i - \vec{\nabla}_j)}{2}$ and $\overleftarrow{k}_{ij} = -\frac{i(\overleftarrow{\nabla}_i - \overleftarrow{\nabla}_j)}{2}$, respectively.

A more detailed description of the theoretical background can be found in Refs. [11,23–25]. In this work we perform spherical Hartree-Fock (HF)-based random-phase approximation (RPA) calculations of strength functions S(E) and consequently determine the centroid energies. The strength function is defined by the sum over all RPA states $|n\rangle$ of energy $E_n$ as

$$S(E) = \sum_n |\langle 0|F_L|n\rangle|^2 \delta(E_n - E_0). \quad (2)$$

In Eq. (2) the electromagnetic single-particle scattering operator, $F_L$, takes the form $F_L = \sum_i f(r_i) Y_{L0}(i)$ or $\frac{Z}{A}\sum_n f(r_n)Y_{L0}(n) - \frac{N}{A}\sum_p f(r_p)Y_{L0}(p)$ for the isoscalar (T=0) or the isovector (T=1) excitations, respectively. The various multipolarities are determined by the operator $f(r)$: for the isoscalar (T = 0) and isovector (T = 1) monopole (L = 0) and quadrupole (L = 2) $f(r) = r^2$, for the octupole (L = 3) $f(r) = r^3$, for the isovector dipole (T = 1, L = 1) $f(r) = r$ and lastly, for the isoscalar dipole (T = 0, L = 1) $f(r) = r^3 - (5/3)\langle r^2\rangle r$ where we have subtracted the contribution from the spurious state [26,27]. The energy moments, $m_k$, of the strength function are integrated within the appropriate energy range $\Omega_1$ to $\Omega_2$:

$$m_k = \int_{\Omega_1}^{\Omega_2} E^k S(E)\, dE \quad (3)$$

from which we determine the centroid energy:

$$E_{CEN} = \frac{m_1}{m_0}. \quad (4)$$

In the RPA calculations we use all the interaction terms from the HF to ensure self-consistency [28]. Similar to previous work in the literature, we adopt the occupation number approximation for the single-particle orbits for the open-shell nuclei to account for the effect of pairing. It is well established that this is a good approximation for the strength functions for excitation energies in the giant resonance regions, resulting in centroid energies within less than 0.3 MeV of the energies obtained by including the effect of pairing, see for example, Ref. [29] and the online documentation of the TDHF code Sky3D [30]. The parameters of each Skyrme interaction used in this work are defined in Table I of reference [11] while the condition for the application of each are presented in Table II of reference [11]. An extended discussion can be found in Section II of Ref. [24]. We setup our calculations of the strength in a box of 100 mesh points with a mesh size of 0.2 fm. The maximum cutoff single particle energy was varied only for the different multipolarities, with 100, 80, 50 and 50 MeV used for multipolarities L = 0, 1, 2 and 3, respectively. The calculations of the energy moments were carried out using $\Gamma = 0.1$ MeV, in the Lorentzian smearing of the strength function S(E), within the energy ranges shown in Table I. The calculated values of $E_{CEN}$ are numerically accurate within 0.1 MeV.

## III. RESULTS

We have calculated the isoscalar and isovector centroid energies, $E_{CEN}$, of multipolarity L=0-3, within the HF-based RPA approximation using 33 Skyrme-type effective interactions of the standard form, for the nuclei $^{44}$Ca, $^{54}$Fe, $^{64,68}$Zn, $^{56,58,60,68}$Ni. The following interactions were employed in the calculations: SGII [31], KDE0 [21], KDE0v1 [21], SKM∗ [32], SK255 [33], SkI3 [34], SkI4 [34], SkI5 [34], SV-bas [35], SV-min [35], SV-sym32 [35], SV-m56-O [36], SV-m64-O [36], SLy4 [37], SLy5 [37], SLy6 [37], SkMP [38], SkO [39], SkO' [39], LNS [40], MSL0 [41], NRAPR [42], SQMC650 [43], SQMC700 [43], SkT1 [44], SkT2 [44], SkT3 [44], SkT8 [44], SkT9 [44], SkT1∗ [44], SkT3∗ [44], Skxs20 [45] and Zσ [46].

We have also calculated the Pearson linear correlation coefficient, C, between the centroid energies, $E_{CEN}$, of each resonance and each of the NM properties: the incompressibility coefficient $K_{NM} = 9\rho_0^2 \frac{\partial^2 E_0}{\partial \rho^2}|_{\rho_0}$, where $E_0[\rho]$ is the binding energy per nucleon and $\rho_0$ is the saturation density,

the effective mass m*/m, the symmetry energy coefficients at $\rho_0$: $J = E_{sym}[\rho_0]$, and its first and second derivatives $L = 3\rho_0 \left.\frac{\partial E_{sym}}{\partial \rho}\right|_{\rho_0}$ and $K_{sym} = 9\rho_0^2 \left.\frac{\partial^2 E_{sym}}{\partial \rho^2}\right|_{\rho_0}$, respectively, $\kappa$, the enhancement coefficient of the energy weighted sum rule (EWSR) of the isovector giant dipole resonance (IVGDR), and $W_0$, the strength of the spin-orbit interaction. We determined the sensitivity of $E_{CEN}$ to bulk properties of NM. For two quantities, x and y, the Pearson linear correlation coefficient is determined by:

$$C = \frac{\sum_{i=1}^{n}(x_i - \bar{x})(y_i - \bar{y})}{\sqrt{\sum_{i=1}^{n}(x_i - \bar{x})^2}\sqrt{\sum_{i=1}^{n}(y_i - \bar{y})^2}} \qquad (4)$$

where $\bar{x}$ and $\bar{y}$ are the averages of x and y and the sum runs over all interactions (n = 33). We adopt the same classification for the degree of correlation as in our previous work [11]: strong ($|C| > 0.80$), medium ($|C| = 0.61 – 0.80$), weak ($|C| = 0.35 – 0.60$) and no correlation ($|C| < 0.35$).

In the sub-sections that follow we discuss the calculated results for giant resonances of multipolarity L = 0-3, beginning with the isoscalar giant resonances and then the isovector giant resonances. We compare our results with the experimental data summarized in Table II. The isotopes of $^{56,68}$Ni are unstable and the data for the isoscalar centroid energies was acquired using inverse kinematics [47–49], whereas for the other isotopes considered here the isoscalar data was acquired at Texas A&M University using inelastic scattering of 240 MeV alpha particles [18–20]. Further details about the experimental setup can be found in [14,50–53]. The experimental data for the isovector giant dipole resonance was taken from the online tabulation maintained by the Centre for Photonuclear Experiments (Moscow State University) [54]. In Table III we show the calculated Pearson linear correlation coefficient between the centroid energies and the various NM properties calculated for each multipolarity.

## A. Isoscalar giant monopole resonance

The centroid energy, $E_{CEN}$, of the isoscalar giant monopole resonance (ISGMR) is plotted against the nuclear matter incompressibility coefficient, $K_{NM}$, of the corresponding interaction used in the HF-based RPA, in FIG. 1. Each nucleus is shown in its own panel, with the calculated values of $E_{CEN}$ shown as full circles and the corresponding experimental data is contained within the dashed lines. We find a medium correlation between the calculated values of $E_{CEN}$ and $K_{NM}$ with C~0.73. For $^{44}$Ca, $^{54}$Fe, $^{56,58}$Ni and $^{64}$Zn we find good agreement between the calculated $E_{CEN}$ and the measured value with the interactions with values of $K_{NM}$ between 200-240 MeV yielding the best overall results. These interactions were found to reproduce the ISGMR $E_{CEN}$ very well across a wide range of masses [11]. We note however, that for $^{68}$Zn, (and for $^{60}$Ni) we find that the calculated value of $E_{CEN}$ of all (most) the interactions are above the experimental results. We point out that in the case of $^{68}$Zn the experimental value of $E_{CEN}$ is much lower than that of other nuclei in the region. Lastly, for the $E_{CEN}$ of $^{68}$Ni we find that the prediction of most of the interactions for $E_{CEN}$ are below the experimental result, requiring a value of $K_{NM}$ above 240 MeV. In FIG. 2 we plot the calculated values of $E_{CEN}$ as a function of the effective mass, $m^*/m$, for which we find no correlation (C ~ -0.26). We also find no correlation between the calculated values of $E_{CEN}$ and the symmetry energy coefficient, $J$, (C~ -0.04), see FIG. 3. Similarly, we don't find any correlation between $E_{CEN}$ and the first derivative of $J$, $L$ (C ~ 0.16) or the second derivative of $J$, $K_{sym}$, shown in FIG. 4, with a Pearson linear correlation coefficient C ~ 0.24. We do not find any correlation with any of the other nuclear matter properties or with $W_0$, see Table III. We note that in our other works with different nuclei, in Refs. [11,12], we found stronger correlations between the values of $E_{CEN}$ of the ISGMR and the value of $K_{NM}$ and of $m^*/m$.

In FIG. 5a we show the values $E_{CEN}$ of the ISGMR of $^{44}$Ca, $^{54}$Fe and $^{64,68}$Zn as functions of the nucleon mass, A. The experimental data are shown by the solid vertical lines while the theoretical values are shown as the dots (connected by lines to guide the eye). As can be seen from these figures most of the interactions predict a slight increase in the value of $E_{CEN}$ going from $^{44}$Ca to $^{54}$Fe and then a steady decrease. The experimental result on the other hand is pretty constant for the first three nuclei and then decreases for $^{68}$Zn. In FIG. 5b we show a similar plot of the $E_{CEN}$ of the ISGMR, but for the Ni isotopes, as a function of A. In the case of the theoretical results we find a steady decrease in the value of $E_{CEN}$ with a kink for the $^{58}$Ni isotope. In the experimental data on the other hand we see that the value of $E_{CEN}$ is similar for $^{56,58}$Ni, decreases slightly for $^{60}$Ni and then increases again for $^{68}$Ni.

### B. Isoscalar giant dipole resonance

In FIG. 6 we show the centroid energy, $E_{CEN}$, of the isoscalar giant dipole resonance (ISGDR), a compression mode, as a function of $K_{NM}$. Each isotope is considered individually, the calculated values are shown as full circles while the experimental data is marked by the dashed lines. We only find a weak correlation between the calculated values of $E_{CEN}$ and $K_{NM}$ (C ~ 0.39) for this compression mode. On the other hand, for the effective mass $m^*/m$ we find a strong correlation with the calculated values of $E_{CEN}$, see FIG. 7 (C ~0.83). Experimental data is not available for $^{54}$Fe and $^{56,68}$Ni. As can be seen from the figure, all calculated values of $E_{CEN}$ are several MeV below the experimental result for $^{44}$Ca and $^{58,60}$Ni. In contrast, the calculated values of $E_{CEN}$, for all the interactions, are 1-4 MeV above the experimental result for $E_{CEN}$ in $^{64,68}$Zn. The experimental values of $E_{CEN}$ for the Zn isotopes are up to 10 MeV below the reported values of the other nuclei considered here. In FIG. 8, we plot the centroid energy of the ISGDR as a function of the symmetry energy coefficient, $J$. We do not find any correlation between the values of $J$ and $E_{CEN}$ (C ~ -0.17). Similarly, for the first and second derivatives of the symmetry energy $J$ ($L$ and $K_{sym}$) we find no correlation, C ~ 0.01 and C ~ 0.23, with the calculated values of $E_{CEN}$, respectively, see Table III.

In FIG. 5c we plot the values of $E_{CEN}$ of the ISGDR of $^{44}$Ca, $^{54}$Fe and $^{64,68}$Zn as functions of their mass, A. We see here that the theory predicts the value of $E_{CEN}$ to gently fluctuate in this region. We also find that theory reproduces the increase in the value of $E_{CEN}$ for the Zn isotopes. However, the calculated $E_{CEN}$ is above the experimental value by a few MeV. In FIG. 5d the $E_{CEN}$ of the Ni isotopes are plotted as a function of A. We see that the theoretical calculations of $E_{CEN}$ are relatively constant across this range of isotopes, while the experimental value of $E_{CEN}$ for $^{58}$Ni is lower than that of $^{60}$Ni. No data is available for the unstable isotopes $^{56,68}$Ni.

## C. Isoscalar giant quadrupole resonance

The calculated centroid energies (shown as full circles) of the isoscalar giant quadrupole resonance (ISGQR) are plotted in FIG. 9 as a function of the effective mass $m^*/m$. The experimental data, available for all the nuclei considered, is marked by the dashed lines. Similar to our previous results for a wide range of nuclear masses [11], we find a strong correlation between the calculated values of $E_{CEN}$ of the ISGQR and $m^*/m$ (C~-0.93). We find that interactions with a lower value of effective mass reproduce the experimental values of $E_{CEN}$ for $^{44}$Ca and $^{54}$Fe the best ($m^*/m$ = 0.65-0.8), whereas for all the other nuclei a higher effective mass ($m^*/m$=0.8-0.9) is in better agreement with the data, with some interactions with effective masses as high as $m^*/m$=1 reproducing the $E_{CEN}$ of $^{56,60}$Ni and $^{64}$Zn. We note that a reasonable agreement between the theoretical and experimental values of $E_{CEN}$ for the considered nuclei is obtained for an interaction associated with the value of $m^*/m$=0.85. In FIG. 10 we plot the $E_{CEN}$ of the ISGQR as a function of the incompressibility coefficient, $K_{NM}$. We find that some interactions across the entire range of $K_{NM}$ seem to reproduce the experimental $E_{CEN}$. Moreover, the correlation between the value of $E_{CEN}$ and $K_{NM}$ is weak (C~0.40), and it is mainly due to the correlation between $K_{NM}$ and $m^*/m$ (see TABLE V of Ref. [11]). As far as the symmetry energy coefficients, $J$, $L$ and $K_{sym}$, we do not find any correlation between the calculated values of $E_{CEN}$ and both $J$ or $L$ (with C~-0.05 and 0.15, respectively), while a weak correlation is obtained between the values of $E_{CEN}$ and $K_{sym}$ (C~0.41). There is also a weak correlation between calculated values of $E_{CEN}$ and the enhancement factor, $\kappa$, of the energy weighted sum rule (EWSR) of the isovector giant dipole resonance (IVGDR) (C~0.52).

In FIG. 5e we plot the values of $E_{CEN}$ of the ISGQR for $^{44}$Ca, $^{54}$Fe and $^{64,68}$Zn as functions of their mass, A. We find that the calculated and the experimental results agree on an increase in the value of $E_{CEN}$ from $^{44}$Ca to $^{54}$Fe and then a decrease for $^{64,68}$Zn. This peculiar behavior was already noticed in our work over a wide range of nuclei in the case of the lighter nuclei considered [11] but not in the region of A=90-100 [12,55]. The value of the $E_{CEN}$ for the Ni isotopes on the other hand, shown in FIG. 5f, seems to steadily decrease in the theoretical calculations but to stay more or less constant in the experimental result. We also reiterate the point made above, which can be clearly seen from FIG. 5e and f, that the interactions with a higher value of effective mass (therefore with a lower value of $E_{CEN}$) reproduce the data of the Zn and Ni isotopes while the experimental values of $E_{CEN}$ for $^{44}$Ca and $^{54}$Fe are reproduced by interactions with a lower value of m*/m.

### D. Isoscalar giant octupole resonance

In FIG. 11 we show a strong correlation between the calculated HF-RPA centroid energy (shown as full circles) of the isoscalar giant octupole resonance (ISGOR) and the effective mass m*/m (C ~ 0.89). As can be seen in the figure the theoretical calculations are above the dashed lines representing the experimental data, available only for $^{58,60}$Ni. A similar result was obtained over a wide range of nuclei [11]. In FIG. 12 we plot the values of $E_{CEN}$ as a function of the incompressibility coefficient for which we find no correlation (C ~ 0.33). Similarly, we find no correlation between $E_{CEN}$ and the symmetry energy coefficients $J$, $L$ and $K_{sym}$, with C ~ -0.1, C ~ -0.01 and 0.23, respectively. We obtained a weak correlation between the calculated values of $E_{CEN}$ and the enhancement factor, $\kappa$, of the EWSR of the IVGDR, C ~ -0.58.

In FIG. 5g we plot the calculated values of $E_{CEN}$ of the ISGOR for $^{44}$Ca, $^{54}$Fe and $^{64,68}$Zn as functions of their mass, A. We find a zig-zag-like trend in this case, a peculiarity not seen in the mass A=90-100 region [12]. For the Ni isotopes, shown in FIG. 5h, we find an overall decrease in the calculated values of $E_{CEN}$ as A increases. Moreover, the experimental values of $E_{CEN}$ for $^{58}$Ni and $^{60}$Ni are significantly lower than the theoretical values, with the $E_{CEN}$ of $^{58}$Ni below that of $^{60}$Ni.

### E. Isovector giant monopole resonance

We point out that one expects that the centroid energy $E_{CEN}$ of the isovector giant monopole resonance (IVGMR), an isovector compression mode, will be sensitive to the incompressibility coefficient, $K_{NM}$, and to the symmetry energy, $J$ and its derivatives, $L$ and $K_{sym}$. In FIG. 13 we show the calculated centroid energies (full circles) of the IVGMR as functions of $K_{NM}$. There is no experimental data for $E_{CEN}$ of the IVGMR for the nuclei studied here. Most of the interactions predict the value of $E_{CEN}$ to be between 28-35 MeV except for $^{54}$Fe and $^{56}$Ni for which several interactions predict a value of $E_{CEN}$ as high as 38.5 MeV. As seen from the figure, we do not find any correlation between the values of $E_{CEN}$ of this isovector compression mode and $K_{NM}$ (C ~ 0.22). In FIG. 14 the calculated $E_{CEN}$ is plotted as a function of the effective mass, $m^*/m$. We obtain a medium correlation between the calculated values of the centroid energy of the IVGMR and $m^*/m$ with a Pearson linear correlation coefficient C ~ -0.64 for the nuclei considered. Next, we plot the calculated $E_{CEN}$ as a function of the symmetry energy coefficient, $J$, in FIG. 15. We find no correlation between the calculated values of $E_{CEN}$ for the IVGMR and $J$ (C ~ -0.24). In FIG. 16, the centroid energies are plotted as a function of the enhancement coefficient, $\kappa$, of the energy weighted sum rule (EWSR) for the isovector giant dipole resonance (IVGDR). The calculated Pearson correlation coefficient between the calculated values of $E_{CEN}$ and $\kappa$ is C ~ 0.80, demonstrating a strong correlation between the two quantitates in agreement with our previous work over a wide range of nuclei [11]. The remaining correlations between the calculated values of $E_{CEN}$ and the various NM properties can be seen in Table III.

In FIG. 17 a, we plot the calculated value of the centroid energy of the IVGMR as a function of mass number A for $^{44}$Ca, $^{54}$Fe and $^{64,68}$Zn. We find that most interactions predict an increase in the value of $E_{CEN}$ when going from $^{44}$Ca to $^{54}$Fe and then lower values for $^{64,68}$Zn. Similarly, we find that many interactions predict a larger value of $E_{CEN}$ for $^{68}$Zn compared to its lighter isotope, $^{64}$Zn, however the two values are very close to each other. In FIG. 17 b we show a similar plot to (a) but for the Ni isotopes. We see a decreasing trend as the mass increases. A particularly steep decrease from the unstable isotope $^{56}$Ni to $^{58}$Ni is obtained for the interactions with a high value of the enhancement coefficient, $\kappa$.

F. Isovector giant dipole resonance

In FIG. 18, the calculated centroid energy, $E_{CEN}$, of the isovector giant dipole resonance (IVGDR) is plotted (full circles) against the symmetry energy coefficient, $J$. We find a weak correlation between the calculated values of $E_{CEN}$ and $J$ (Pearson linear correlation coefficient C ~ -0.39). Similarly, in FIG. 19 we plot the calculated $E_{CEN}$ as a function of the enhancement coefficient, $\kappa$, of the EWSR for the IVGDR. We find a strong correlation between the calculated values of $E_{CEN}$ and $\kappa$ (Pearson linear correlation coefficient C ~ 0.80). As shown in FIG. 19 we find that the values of $E_{CEN}$ of most of the interactions fall below the experimental data for $^{44}$Ca with only the $E_{CEN}$ of the interactions with a higher value of $\kappa$ (> 0.6) coming close to the experimental result, and similarly for $^{56,60}$Ni and $^{64}$Zn. We obtained the opposite result for $^{54}$Fe and $^{68}$Zn, where we find that the experimental data for $E_{CEN}$ is also reproduced by interactions with a smaller enhancement coefficient $\kappa$ as low as 0.1. In $^{58,68}$Ni we obtain good agreement between theory and experiment for interactions with a value of $\kappa$ = 0.25-0.7. Overall, we find that interactions with a value of the enhancement coefficient $\kappa$ = 0.25-0.7 are the best at reproducing all the nuclei considered, in agreement with our study over a wide range of masses [11]. In FIG. 20 we plot the $E_{CEN}$ as a function of the effective mass, m*/m. We obtain a medium correlation between the values of $E_{CEN}$ and m*/m with a Pearson linear correlation coefficient C ~ -0.62. The correlation coefficients for the remaining NM properties are shown in Table III.

In FIG. 17 c, the values $E_{CEN}$ of the IVGDR are plotted as functions of the mass number A of the isotopes of $^{44}$Ca, $^{54}$Fe and $^{64,68}$Zn. We find that for these nuclei the calculated value of the centroid energy decreases smoothly as A increases, with some minor fluctuations for some interactions. The experimental result on the other hand shows some fluctuation with the $E_{CEN}$ of $^{64}$Zn higher than the $E_{CEN}$ of $^{54}$Fe. The centroid energy of the Ni isotopes is plotted as a function of A in FIG. 17 d. We find that the theory predicts the centroid energy of the unstable isotope $^{56}$Ni to be lower than that of $^{58}$Ni, then a smooth decrease in the value of $E_{CEN}$ as A increases to 60 and 68. Experimentally however, the centroid energy of $^{58}$Ni is below that of $^{56}$Ni but roughly the same as $^{60}$Ni, while $^{68}$Ni is lower.

### G. Isovector giant quadrupole resonance

We consider now the isovector giant quadrupole resonance (IVGQR). In this case no experimental data for the centroid energy $E_{CEN}$ is available for the nuclei consider here. We plot the calculated values of the centroid energy $E_{CEN}$ (full circles) as a function of the symmetry energy coefficient, $J$, in FIG. 21 We find that most of the calculated centroid energies of the isotopes considered fall between 24-35 MeV. We find a weak correlation between the calculated values of $E_{CEN}$ and J with a calculated Pearson linear correlation coefficient C ~ -0.38. We do not find any correlation between the calculated values of $E_{CEN}$ and the first or second derivative of $J$, $L$ and $K_{sym}$ with C ~ -0.34 and C ~ -0.17, respectively. In FIG. 22 we show the calculated centroid energy as a function of the enhancement coefficient of the EWSR of the IVGDR, $\kappa$. In this case we find a strong correlation (C ~ 0.81) between the two values. The calculated centroid energies are plotted as a function of effective mass in FIG. 23 and we obtained a medium correlation between the two values (C ~ -0.73). As shown in Table III, no correlation is found between the calculated values of $E_{CEN}$ and the incompressibility coefficient $K_{NM}$.

In FIG. 17 e, the calculated $E_{CEN}$ of the IVGQR is plotted as a function of the mass number A for the isotopes of $^{44}$Ca, $^{54}$Fe and $^{64,68}$Zn. In this case we find that the centroid energy for all 4 isotopes is predicted to decrease slowly as A increases. Similarly, in FIG. 17f we plot the $E_{CEN}$ for the Ni isotopes as a function of mass A and find that for most interactions the value of the centroid energy slowly decreases as A is increasing.

### H. Isovector giant octupole resonance

Experimental data for the centroid energy, $E_{CEN}$, of the isovector giant octupole resonance (IVGOR) is unavailable for the nuclei consider here. We show the calculated centroid energy (full circles) as a function of the symmetry energy coefficient $J$ in FIG. 24. We find that the calculated values of centroid energy associated with most of the interactions considered are in the range 34-43MeV. No correlation is found between the calculated values of $E_{CEN}$ and the symmetry energy coefficient $J$, with a Pearson linear correlation coefficient of C ~ -0.29. Similarly, for the first and second derivative of $J$ ($L$ and $K_{sym}$) we obtained the values of C ~ -0.18 and C ~ 0.01, respectively. In FIG. 25 we plot the calculated centroid energies as a function of the enhancement coefficient $\kappa$ of the EWSR for the IVGDR. As can be seen from the figure we obtained a medium correlation between the calculated values of $E_{CEN}$ and $\kappa$ with a Pearson linear correlation coefficient of C ~ 0.79. In FIG. 26 we show the calculated values of $E_{CEN}$ as a function of the effective mass m*/m. We determined a strong correlation between the calculated values of $E_{CEN}$ and m*/m with a Pearson linear correlation coefficient of C ~ -0.82. We do not find any correlation between the values of $E_{CEN}$ and $K_{NM}$ (C ~ 0.23).

The calculated values of $E_{CEN}$ for the IVGOR are plotted as functions of nuclear mass number A for $^{44}$Ca, $^{54}$Fe and $^{64,68}$Zn in FIG. 17g while the Ni isotopes are shown in FIG. 17h. In both cases we find a slow decrease in the value of $E_{CEN}$ as A increases. However, for the case of the Ni isotopes, we find that some interactions predict the $E_{CEN}$ of $^{58}$Ni to be above that of the unstable isotope $^{56}$Ni.

## IV. SUMMARY AND CONCLUSIONS

We have carried out fully self-consistent spherical HF-RPA calculations, using the occupation number approximation for open shells, for 33 Skyrme-type effective nucleon-nucleon interaction, and obtained the centroid energies, $E_{CEN}$, of both isoscalar and isovector giant resonances of multipolarities L = 0-3 for the isotopes of $^{44}$Ca, $^{54}$Fe, $^{64,68}$Zn and $^{56,58,60,68}$Ni. We compared our isoscalar results with the data recently obtained at Texas A&M University [18,19] and the $^{56,68}$Ni experiments [20], while the isovector data was taken from the online 'Centre for photonuclear experimental data" maintained by Moscow State University [54]. It is important to point out when comparing the theoretical prediction of the 33 Skyrme interaction with experimental data that we have encountered important disagreements. In particular:

(i) For the isoscalar giant monopole resonance (ISGMR) we found for $^{44}$Ca, $^{54}$Fe, $^{56,58}$Ni and $^{64}$Zn good agreement between the calculated $E_{CEN}$ and the measured values with the interactions associated with values of $K_{NM}$ between 200-240 MeV. We note however, that for $^{68}$Zn, (and for $^{60}$Ni) the calculated values of $E_{CEN}$ of all (most) the interactions are above the experimental results. The experimental value of $E_{CEN}$ for $^{68}$Zn is much lower than that of other nuclei in the region. Lastly, for $^{68}$Ni the prediction of most of the interactions for $E_{CEN}$ are below the experimental result, requiring a value of $K_{NM}$ above 240 MeV.

(ii) For the isoscalar giant dipole resonance (ISGDR) and the isoscalar giant octupole resonance (ISGOR), there is significant disagreement between theory and experiment. Surprisingly, for $^{44}$Ca, $^{58,60}$Ni the calculated centroid energies are significantly lower than the experimental result, opposite to what we found in other nuclei studied [11,12]. The experimental result for the ISGOR centroid energy for $^{58,60}$Ni on the other hand are significantly lower than the calculated values.

(iii) For the isoscalar giant quadruple resonance (ISGQR), we found that interactions with a lower value of effective mass reproduce the experimental values of $E_{CEN}$ for $^{44}$Ca and $^{54}$Fe the best (m*/m = 0.65-0.8), whereas for all the other nuclei a higher effective mass (m*/m=0.8-0.9) is in better agreement with the data with some interactions with effective masses as high as m*/m=1.0 reproducing the $E_{CEN}$ of $^{56,58,60}$Ni and $^{64}$Zn. We note that a reasonable agreement between the theoretical and experimental values of $E_{CEN}$ for the considered nuclei is obtained for an interaction associated with the value of m*/m = 0.85.

(iv) For the isovector giant dipole resonance (IVGDR), we found that the values of $E_{CEN}$ of most of the interactions fall below the experimental data for $^{44}$Ca with only the $E_{CEN}$ of the interactions with a higher value of κ (> 0.6) coming close to the experimental result. We obtained the opposite result for $^{54}$Fe and $^{68}$Zn, where we found that the experimental data for $E_{CEN}$ is also reproduced by interactions with a smaller enhancement coefficient κ, as low as low as 0.1. Overall, we found that interactions with a value of the enhancement coefficient κ = 0.25-0.7 are the best at reproducing all the nuclei considered, in agreement with our study over a wide range of masses [11].

We determined the Pearson linear correlation coefficient, C, in order to quantify the sensitivity of the calculated centroid energies of the giant resonance to the nuclear matter properties associated with the adopted Skyrme-type effective interactions given in Eq. (1). For the correlations between the calculated values of $E_{CEN}$ and the nuclear matter incompressibility coefficient, $K_{NM}$, we found medium, weak and no correlations for the compression modes of the ISGMR, ISGDR and isovector giant monopole resonance (IVGMR), respectively. For the correlations between the calculated values of $E_{CEN}$ and the effective mass, m*/m, we found strong correlations for the ISGDR, ISGQR, ISGOR and isovector giant octuple resonance (IVGOR) and medium correlations for the IVGMR, IVGDR and isovector giant quadrupole resonance (IVGQR). We also found strong correlations between the calculated values of $E_{CEN}$ and the enhancement coefficient, κ, of the EWSR of the IVGDR, for the IVGMR, IVGDR and IVGQR while medium correlations is found for the IVGOR. For the symmetry energy coefficients *J*, and its first derivative *L* and second derivative $K_{sym}$ we found at most weak correlations with the calculated $E_{CEN}$ values of certain multipolarities. We note that we found slightly lower values for C in the set of nuclei considered in this work compared to the mostly larger mass region of nuclei we studied recently [11,12]. We note that for some of the nuclei considered in this work several different configurations (set of occupation numbers) were feasible, an effect not seen for the other nuclei we recently studied [11,12].

The significant disagreement between the theoretical and the experimental values of $E_{CEN}$ for the ISGDR (particularly $^{44}$Ca, $^{58,60}$Ni) and ISGOR (data available only for $^{58,60}$Ni) as well as marginal agreement in other cases (such as the ISGMR in $^{68}$Zn) suggests further investigations could be useful. The calculations of the response functions could be carried beyond the mean-field approximation by including nuclear structure effects [56–59], while parametrized ground state density and semi-classical transition densities that were used in the analyses of the experimental data within the folding-model distorted wave Born approximation could be replaced with calculated HF-based RPA ground state and transition densities [26,60].

## ACKNOWLEDGEMENT

This work was supported in part by the US Department of Energy under Grant No. DE-FG03-93ER40773.

# Figure Captions

FIG. 1. Calculated centroid energies, $E_{CEN}$, in MeV (full circle) of the isoscalar giant monopole resonance (ISGMR), for different Skyrme interactions, as a function of the incompressibility coefficient $K_{NM}$. Each nucleus has its own panel, the experimental uncertainties are contained by the dashed lines. We find a medium correlation between the values of $E_{CEN}$ and $K_{NM}$ with a Pearson linear correlation coefficient C ~ 0.73.

FIG. 2. Similar to FIG. 1 for the effective mass $m^*/m$. We find no correlation between the calculated values of $E_{CEN}$ and $m^*/m$ with a Pearson linear correlation coefficient C ~ -0.26.

FIG. 3. Similar to FIG. 1 for the symmetry energy $J$ at saturation energy. We do not find any correlation between the calculated values of $J$ and $E_{CEN}$, with a Pearson linear correlation coefficient C ~ -0.04.

FIG. 4. Similar to FIG. 1 for the second derivative of the symmetry energy, $K_{sym}$, at saturation energy. We don't find any correlation between the calculated values of $K_{sym}$ and $E_{CEN}$, with a Pearson linear correlation coefficient C ~ 0.24.

FIG. 5. The centroid energies [MeV] of the isoscalar giant resonances of multipolarities L=0-3 for $^{44}$Ca, $^{54}$Fe, and $^{64,68}$Zn (left figures) and for $^{56,58,60,68}$Ni (right figures), are plotted against the mass A of each isotope. The experimental error bars (where available) are shown by the solid vertical lines, while the theoretical values of $E_{CEN}$ are shown as dots connected by lines (to guide the eye).

FIG. 6. Similar to FIG. 1 for the isoscalar giant dipole resonance (ISGDR) as a function of the incompressibility coefficient $K_{NM}$, for different nuclei. We find weak correlation between the calculated values of $E_{CEN}$ and $K_{NM}$ with a Pearson linear correlation coefficient C ~ 0.39, mostly due to the already recognized correlation between $K_{NM}$ and $m^*/m$ shown in Table V of Ref. [11].

FIG. 7. Similar to FIG. 1 for the ISGDR as a function of $m^*/m$ for different nuclei. We find strong correlation between the calculated values of $E_{CEN}$ and $m^*/m$ with a Pearson linear correlation coefficient C ~ -0.83 in all cases.

FIG. 8. Similar to FIG. 1, for the ISGDR as a function of the symmetry energy $J$ at saturation density for different nuclei. We find no correlation between the calculated values of $E_{CEN}$ and $J$ (Pearson linear correlation coefficient C ~ -0.17).

FIG. 9. Similar to FIG. 1, for the ISGQR as a function of the effective mass $m^*/m$. We find strong correlation between the calculated values of $m^*/m$ and $E_{CEN}$ with a Pearson linear correlation coefficient C ~ -0.93 in all cases.

FIG. 10. Similar to FIG. 1, for the ISGQR as a function of the incompressibility coefficient $K_{NM}$. We find weak correlation between the calculated values of $K_{NM}$ and $E_{CEN}$ with a Pearson linear correlation coefficient close to C ~ 0.40 for all isotopes, mostly due to the correlation between $K_{NM}$ and $m^*/m$ shown in Table V of Ref. [11].

FIG. 11. Similar to FIG. 1, for the ISGOR plotted against the effective mass $m^*/m$. We find strong correlation between the calculated values of $m^*/m$ and $E_{CEN}$ with a Pearson linear correlation coefficient C greater in magnitude than -0.89 in all cases.

FIG. 12. Similar to FIG. 1, for the ISGOR plotted against the incompressibility coefficient $K_{NM}$. We find no correlation between the calculated values of $K_{NM}$ and $E_{CEN}$ with a Pearson linear correlation coefficient roughly C ~ 0.32 in all cases.

FIG. 13. Similar to FIG. 1, for the IVGMR plotted against the incompressibility coefficient of nuclear matter $K_{NM}$. We do not find any correlation between the calculated values of $E_{CEN}$ and $K_{NM}$ with a Pearson linear correlation coefficient C ~ 0.22.

FIG. 14. Similar to FIG. 1, for the IVGMR plotted against the effective mass $m^*/m$. We find medium correlation between the calculated values of $E_{CEN}$ and $m^*/m$ with a Pearson linear correlation coefficient C ~ -0.64 for all the isotopes considered here.

FIG. 15. Similar to FIG. 1, for the IVGMR as a function of the symmetry coefficient $J$ at saturation density. We do not find any correlation between the value of $J$ and $E_{CEN}$ with a Pearson linear correlation coefficient C ~ -0.24.

FIG. 16. Similar to FIG. 1, for the IVGMR plotted against the enhancement coefficient, $\kappa$, of the energy weighted sum rule (EWSR) of the isovector giant dipole resonance (IVGDR). We find strong correlation between the calculated value of $\kappa$ and $E_{CEN}$ with a Pearson linear correlation coefficient C ~ 0.80 for all isotopes considered.

FIG. 17. The centroid energies [MeV] of the isovector giant resonances of multipolarities L=0-3 for $^{44}$Ca, $^{54}$Fe, and $^{64,68}$Zn (left figures) and for $^{56,58,60,68}$Ni (right figures), are plotted against the mass A of each isotope. The experimental error bar (where available) are shown by the solid vertical lines, while the theoretical values of $E_{CEN}$ are shown as dots connected by lines (to guide the eye).

FIG. 18. Similar to FIG. 1, for the IVGDR as a function of the symmetry energy $J$ at saturation density. We find weak correlation between the value of $J$ and the value of $E_{CEN}$ with a Pearson linear correlation coefficient C ~ -0.39.

FIG. 19. Similar to FIG. 1, for the IVGDR for different nuclei plotted against the energy weighted sum rule enhancement coefficient $\kappa$ of the IVGDR. We find strong correlation between the calculated values of $\kappa$ and $E_{CEN}$ with a Pearson linear correlation coefficient C ~ 0.80 for all isotopes considered.

FIG. 20. Similar to FIG. 1, for the IVGDR plotted against the effective mass $m^*/m$. We find medium correlation between the calculated values of $E_{CEN}$ and $m^*/m$ with a Pearson linear correlation coefficient close to C ~ -0.62 for all the isotopes considered here.

FIG. 21. Similar to FIG. 1, for the IVGQR as a function of the symmetry energy $J$ at saturation density. We find a weak correlation between the value of $J$ and the value of $E_{CEN}$, with a Pearson linear correlation coefficient of C ~ -0.38.

FIG. 22. Similar to FIG. 1, for the IVGQR plotted against the enhancement coefficient $\kappa$ of the EWSR of the IVGDR. We find a strong correlation between the calculated values of $\kappa$ and $E_{CEN}$ with a Pearson linear correlation coefficient C ~ 0.81 for all isotopes considered.

FIG. 23. Similar to FIG. 1, for the IVGQR of different nuclei plotted against the effective mass $m^*/m$. We find medium correlation between the calculated values of $E_{CEN}$ and $m^*/m$, with a Pearson linear correlation coefficient close to C ~ -0.73 for all the isotopes considered here.

FIG. 24. Similar to FIG. 1, for the IVGOR as a function of the symmetry energy $J$ at saturation density. We do not find any correlation between the value of $J$ and the value of $E_{CEN}$ with a Pearson linear correlation coefficient C ~ -0.29.

FIG. 25. Similar to FIG. 1, for the IVGOR plotted against the enhancement coefficient $\kappa$ of the EWSR of the IVGDR. We find a medium correlation between the values of $\kappa$ and $E_{CEN}$ with a Pearson linear correlation coefficient C ~ 0.79 for all isotopes considered.

FIG. 26. Similar to FIG. 1, for the IVGOR plotted against the effective mass $m^*/m$. We find strong correlation between the calculated values of $m^*/m$ and $E_{CEN}$ with a Pearson linear correlation coefficient C ~ -0.82 for all isotopes considered.

Table I. Excitation energy ranges in MeV) used for calculating the centroid energies of the isoscalar and isovector giant resonances from the corresponding strength functions.

|      | $^{44}$Ca | $^{54}$Fe | $^{56}$Ni | $^{58}$Ni | $^{60}$Ni | $^{64}$Zn | $^{68}$Zn | $^{68}$Ni |
|------|-----------|-----------|-----------|-----------|-----------|-----------|-----------|-----------|
| **L0T0** | 9 - 40 | 9 - 40 | 12 - 35 | 9 - 40 | 9 - 40 | 9 - 40 | 9 - 40 | 12 - 30 |
| **L1T0** | 20 - 40 | 20 - 40 | 20 - 40 | 20 - 40 | 20 - 40 | 20 - 40 | 20 - 40 | 20 - 40 |
| **L2T0** | 9 - 40 | 9 - 40 | 12 - 35 | 9 - 40 | 9 - 40 | 9 - 40 | 9 - 40 | 12 - 30 |
| **L3T0** | 15 - 40 | 15 - 40 | 15 - 40 | 15 - 40 | 15 - 40 | 15 - 40 | 15 - 40 | 15 - 40 |
| **L0T1** | 7 - 60 | 7 - 60 | 7 - 60 | 7 - 60 | 7 - 60 | 7 - 60 | 7 - 60 | 7 - 60 |
| **L1T1** | 0 - 60 | 0 - 60 | 0 - 60 | 0 - 60 | 0 - 60 | 0 - 60 | 0 - 60 | 0 - 60 |
| **L2T1** | 7 - 60 | 7 - 60 | 7 - 60 | 7 - 60 | 7 - 60 | 7 - 60 | 7 - 60 | 7 - 60 |
| **L3T1** | 25 - 60 | 25 - 60 | 25 - 60 | 25 - 60 | 25 - 60 | 25 - 60 | 25 - 60 | 25 - 60 |

Table II. Experimental values for the centroid energies (in MeV) of isoscalar and isovector giant resonances. The isoscalar data was taken from the following: Ref. [18] for a, Ref. [19] for b, Ref. [47] for c, Ref. [20] for d, Ref. [48] for e, and Ref. [49] for f. The isovector data was taken from the online 'Centre for photonuclear experimental data" maintained by Moscow State University [54].

|  | $^{44}$Ca | $^{54}$Fe | $^{56}$Ni | $^{58}$Ni | $^{60}$Ni | $^{64}$Zn | $^{68}$Zn | $^{68}$Ni |
|---|---|---|---|---|---|---|---|---|
| **L0T0** | 19.49 (34) a | 19.66 (37) b | 19.30 (50) c | 19.32 (32) d | 18.10 (29) d | 18.88 (79) b | 16.60 (17) b | 21.1 (19) e |
| **L1T0** | 35.03 (145) a | - | - | 34.06 (30) d | 36.12 (28) d | 25.66 (121) b | 27.65 (39) b | - |
| **L2T0** | 17.21 (48) a | 18.05 (87) b | 16.20 (50) c | 16.34 (13) d | 15.88 (14) d | 15.85 (31) b | 15.54 (32) b | 15.9 (13) f |
| **L3T0** | - | - | - | 23.20 (30) d | 24.40 (26) d | - | - | - |
| **L1T1** | 21.63 (50) | 18.94 (50) | 20.91 (50) | 20.41 (50) | 20.41 (50) | 19.53 (50) | 17.18 (50) | 17.10 (20) |

Table III. Pearson linear correlation coefficients, C, between the calculated centroid energy of each giant resonance and each nuclear matter property at saturation density.

|  | $K_{NM}$ | $m^*/m$ | $W_0(X_W=1)$ | $J$ | $L$ | $K_{sym}$ | $\kappa$ |
|---|---|---|---|---|---|---|---|
| ISGMR | 0.73 | -0.26 | -0.07 | -0.04 | 0.16 | 0.24 | 0.02 |
| ISGDR | 0.39 | -0.83 | -0.02 | -0.17 | 0.01 | 0.23 | 0.58 |
| ISGQR | 0.40 | -0.93 | 0.07 | -0.05 | 0.15 | 0.41 | 0.53 |
| ISGOR | 0.32 | -0.89 | 0.04 | -0.15 | -0.01 | 0.24 | 0.58 |
| IVGMR | 0.22 | -0.64 | -0.12 | -0.24 | -0.13 | -0.03 | 0.80 |
| IVGDR | 0.09 | -0.62 | -0.12 | -0.39 | -0.40 | -0.27 | 0.80 |
| IVGQR | 0.17 | -0.73 | -0.13 | -0.38 | -0.34 | -0.17 | 0.81 |
| IVGOR | 0.23 | -0.82 | -0.04 | -0.29 | -0.18 | 0.01 | 0.79 |

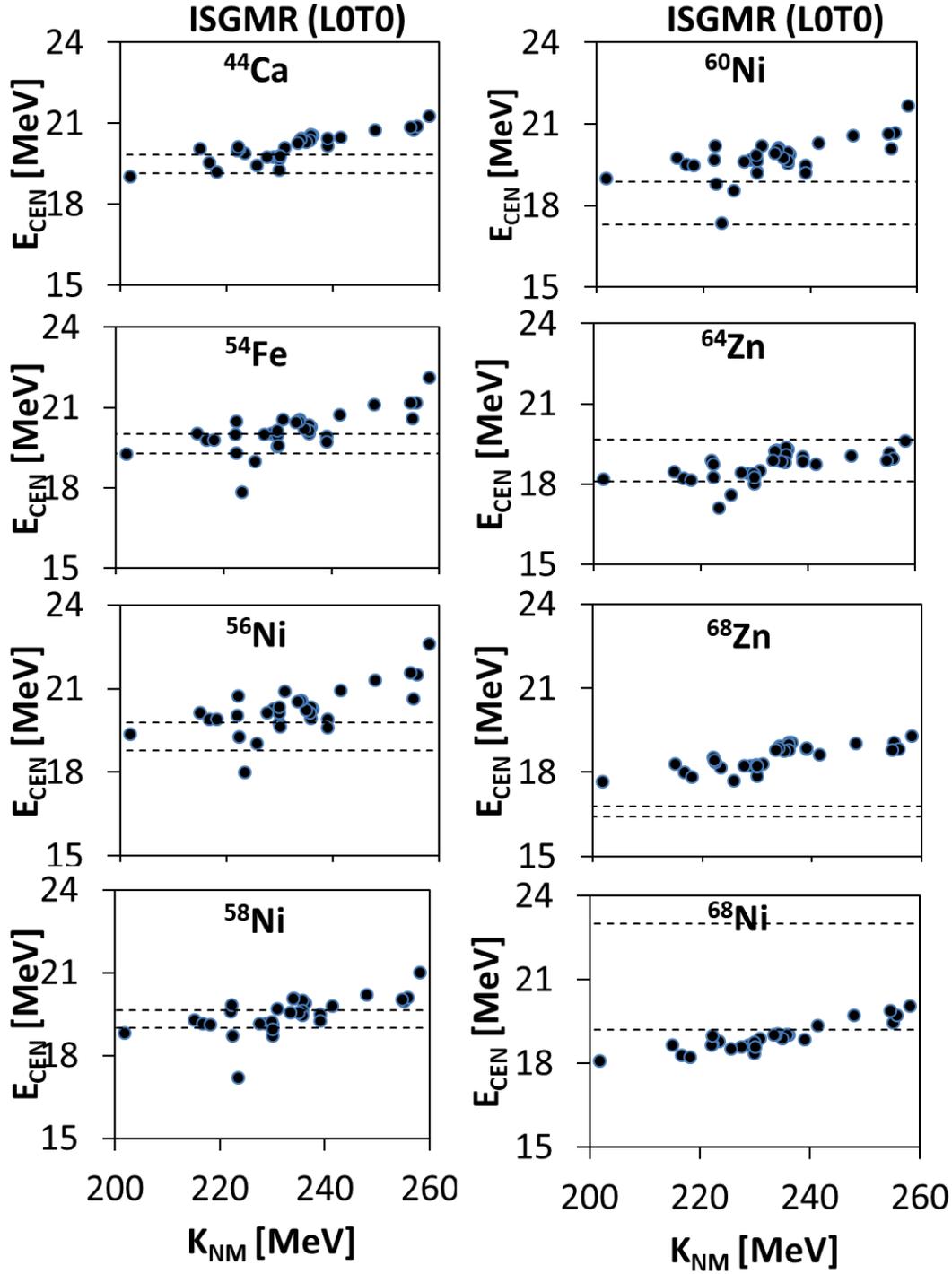

FIG. 1. Calculated centroid energies, $E_{CEN}$, in MeV (full circle) of the isoscalar giant monopole resonance (ISGMR), for different Skyrme interactions, as a function of the incompressibility coefficient $K_{NM}$. Each nucleus has its own panel, the experimental uncertainties are contained by the dashed lines. We find a medium correlation between the values of $E_{CEN}$ and $K_{NM}$ with a Pearson linear correlation coefficient C ~ 0.73.

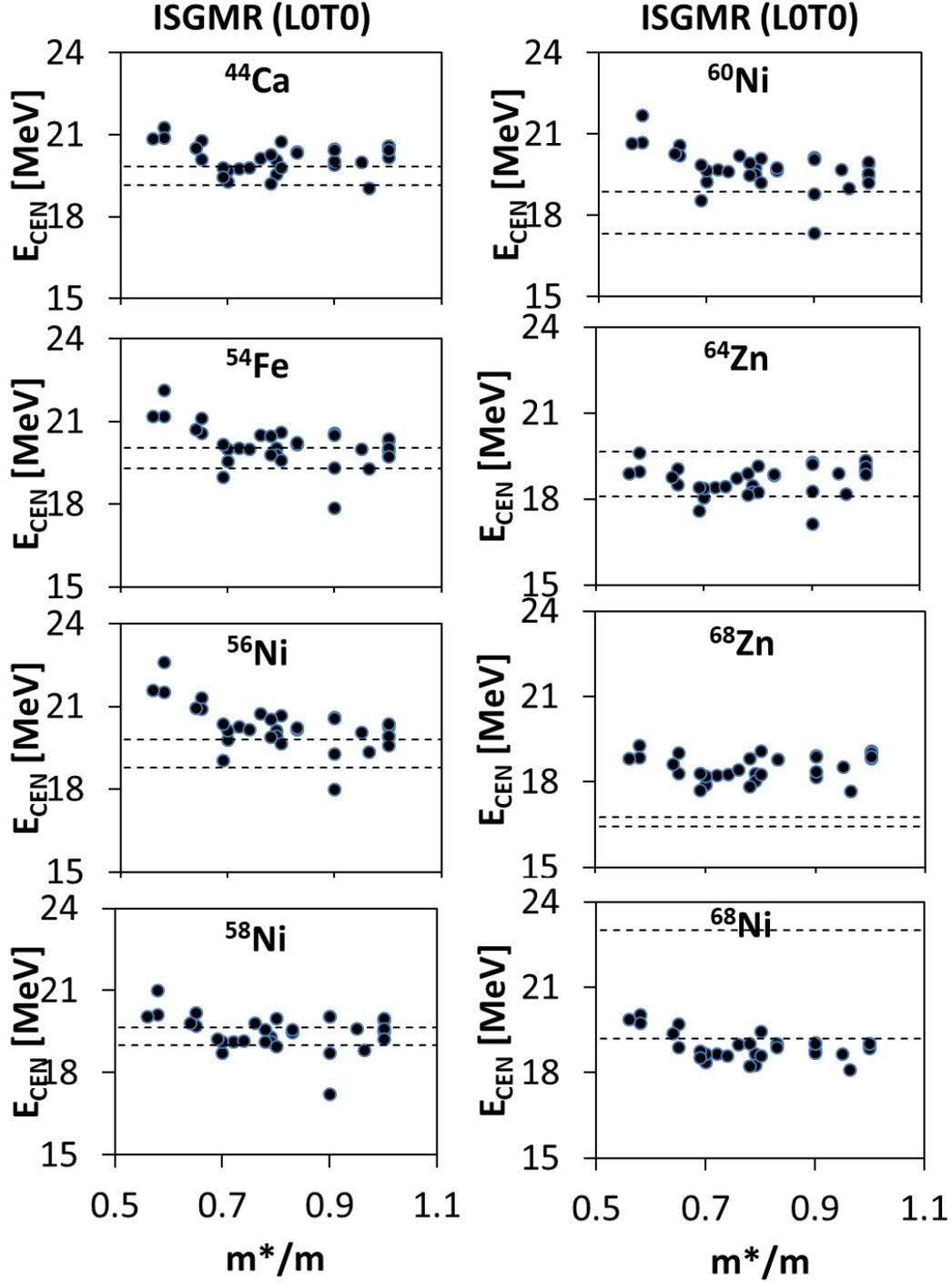

FIG. 2. Similar to FIG. 1 for the effective mass m*/m. We find no correlation between the calculated values of $E_{CEN}$ and m*/m with a Pearson linear correlation coefficient C ~ -0.26.

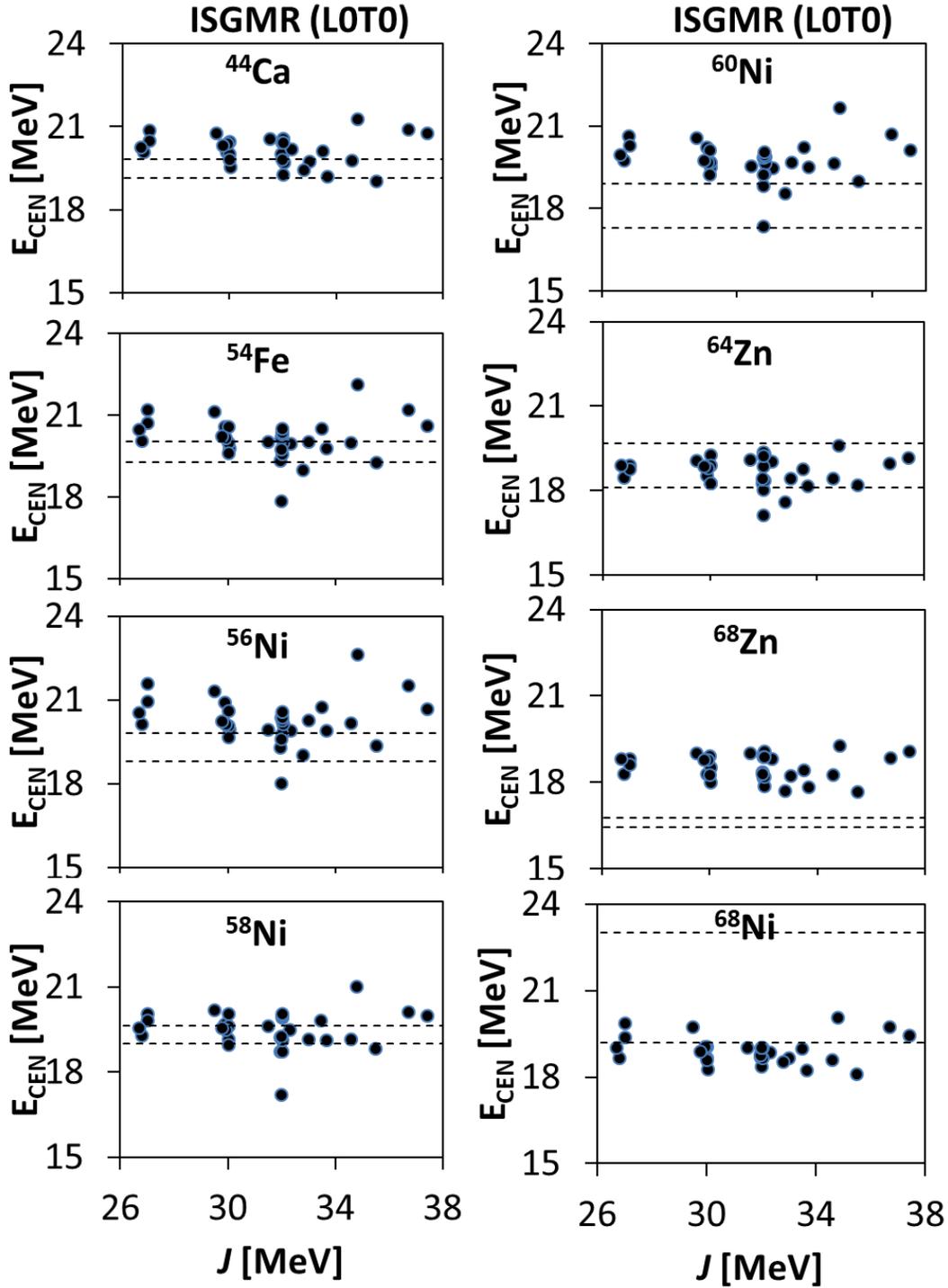

FIG. 3. Similar to FIG. 1 for the symmetry energy $J$ at saturation energy. We do not find any correlation between the calculated values of $J$ and $E_{CEN}$, with a Pearson linear correlation coefficient C ~ -0.04.

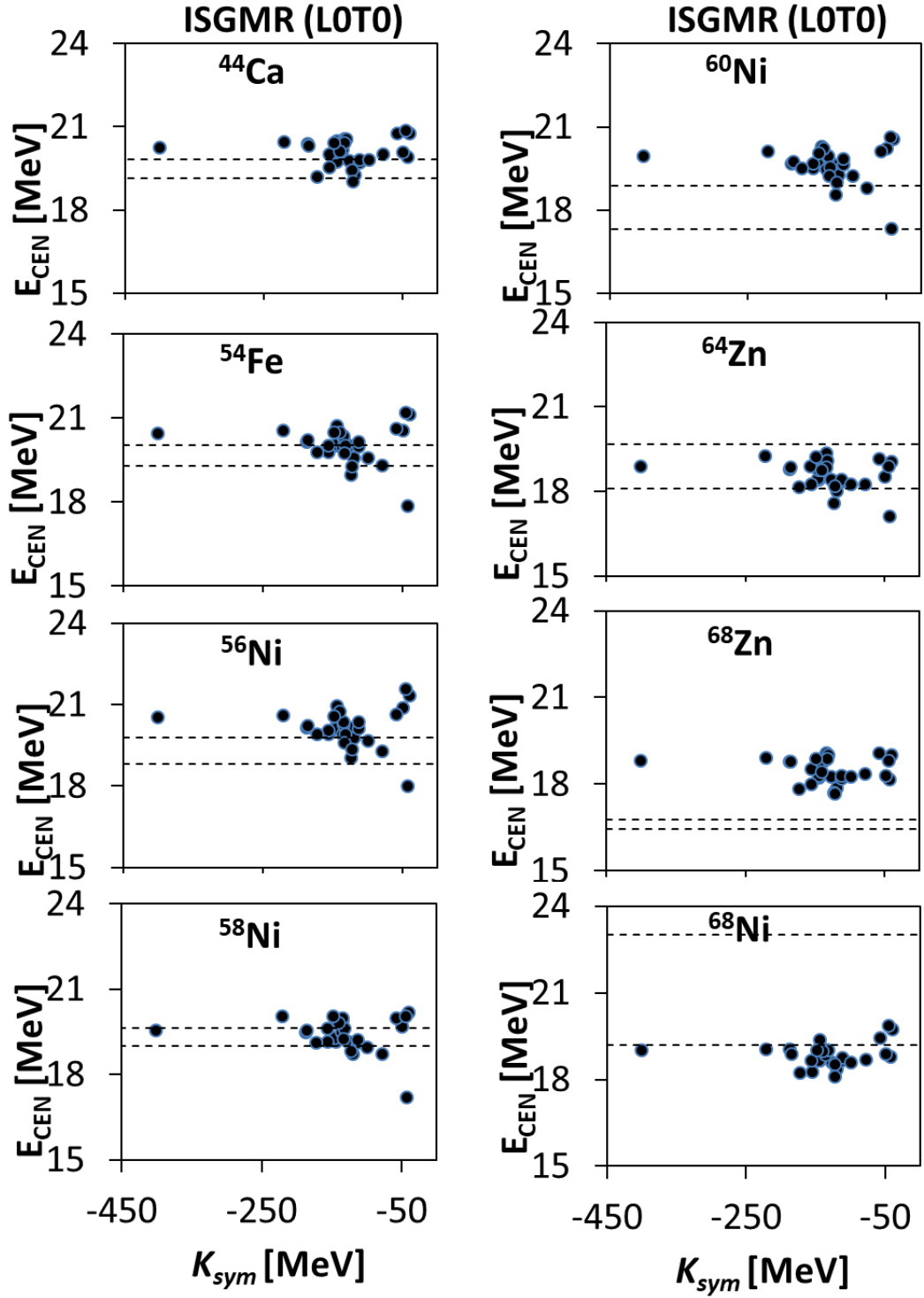

FIG. 4. Similar to FIG. 1 for the second derivative of the symmetry energy, $K_{sym}$, at saturation energy. We don't find any correlation between the calculated values of $K_{sym}$ and $E_{CEN}$, with a Pearson linear correlation coefficient C ~ 0.24.

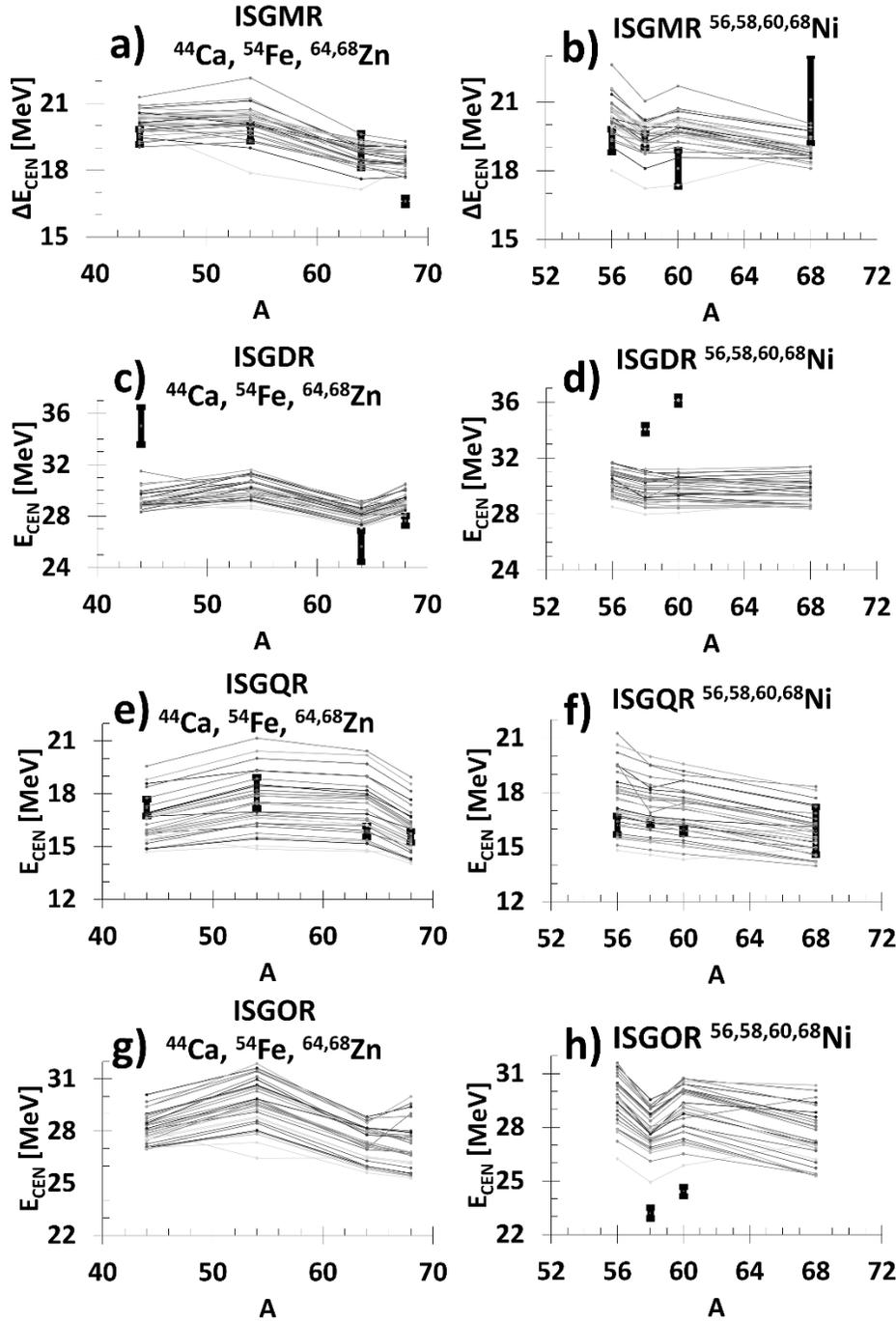

FIG. 5. The centroid energies [MeV] of the isoscalar giant resonances of multipolarities L=0-3 for $^{44}$Ca, $^{54}$Fe, and $^{64,68}$Zn (left figures) and for $^{56,58,60,68}$Ni (right figures), are plotted against the mass A of each isotope. The experimental error bars (where available) are shown by the solid vertical lines, while the theoretical values of $E_{CEN}$ are shown as dots connected by lines (to guide the eye).

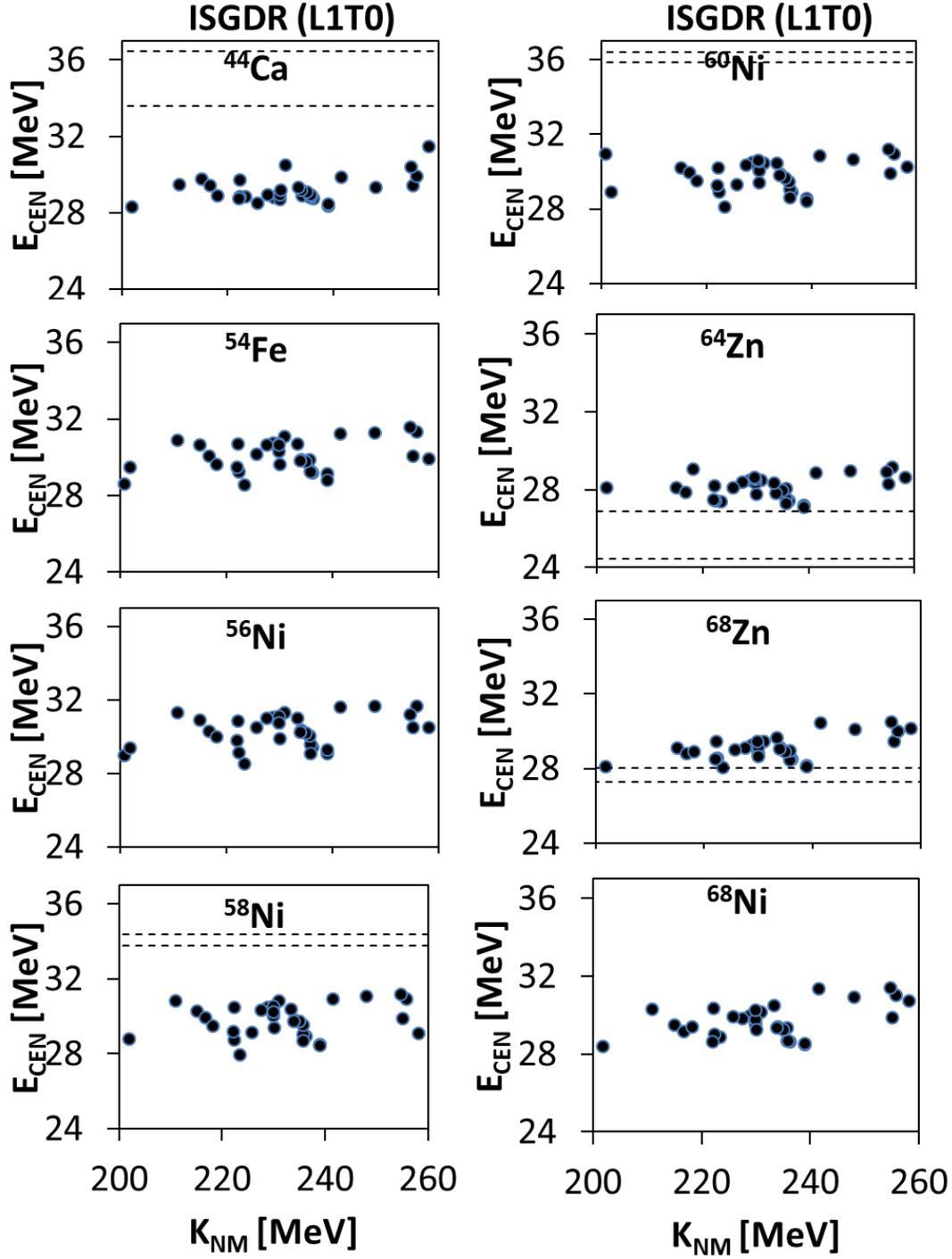

FIG. 6. Similar to FIG. 1 for the isoscalar giant dipole resonance (ISGDR) as a function of the incompressibility coefficient $K_{NM}$, for different nuclei. We find weak correlation between the calculated values of $E_{CEN}$ and $K_{NM}$ with a Pearson linear correlation coefficient $C \sim 0.39$, mostly due to the already recognized correlation between $K_{NM}$ and $m^*/m$ shown in Table V of Ref. [11].

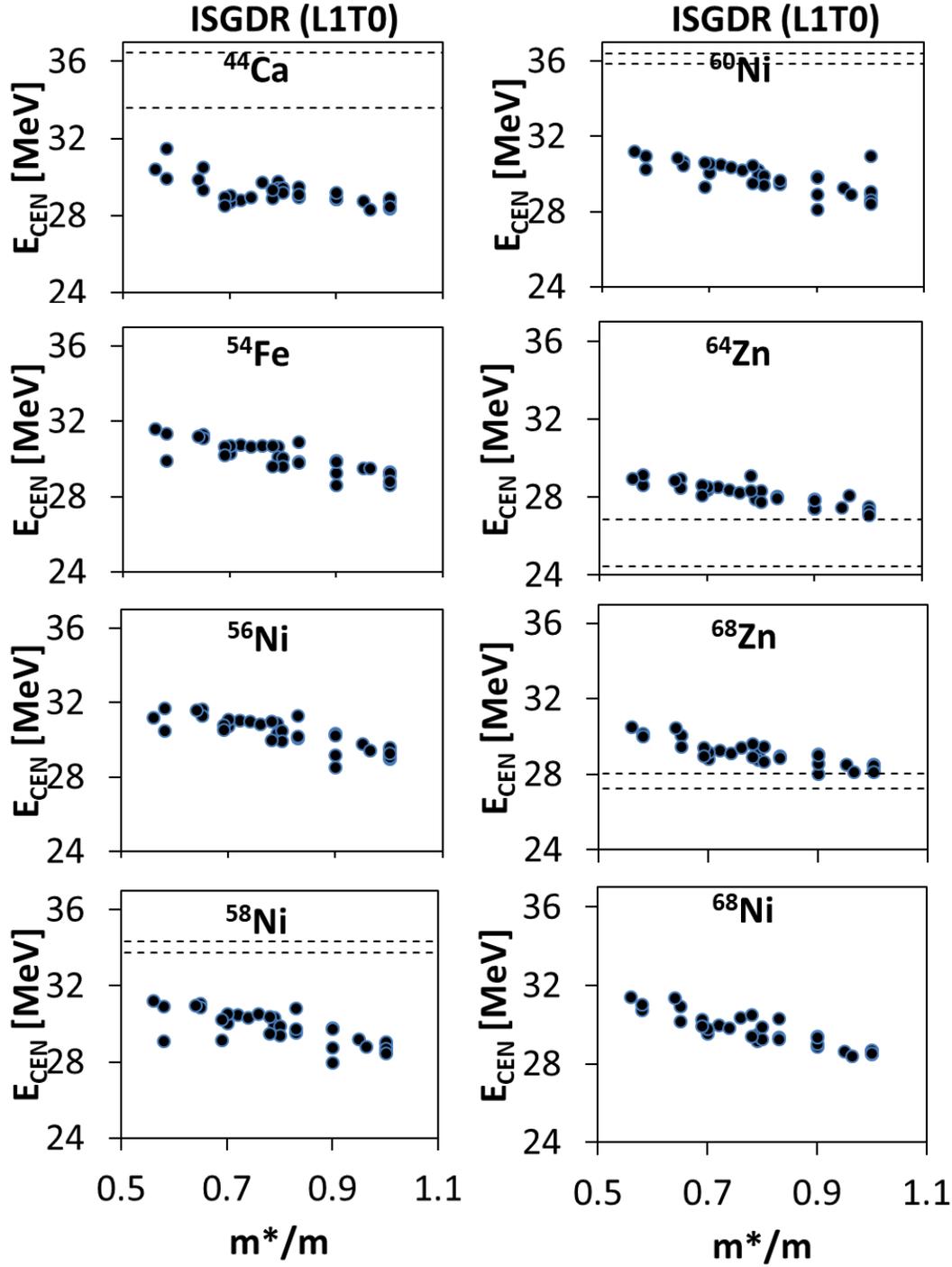

FIG. 7. Similar to FIG. 1 for the ISGDR as a function of $m^*/m$ for different nuclei. We find strong correlation between the calculated values of $E_{CEN}$ and $m^*/m$ with a Pearson linear correlation coefficient $C \sim -0.83$ in all cases.

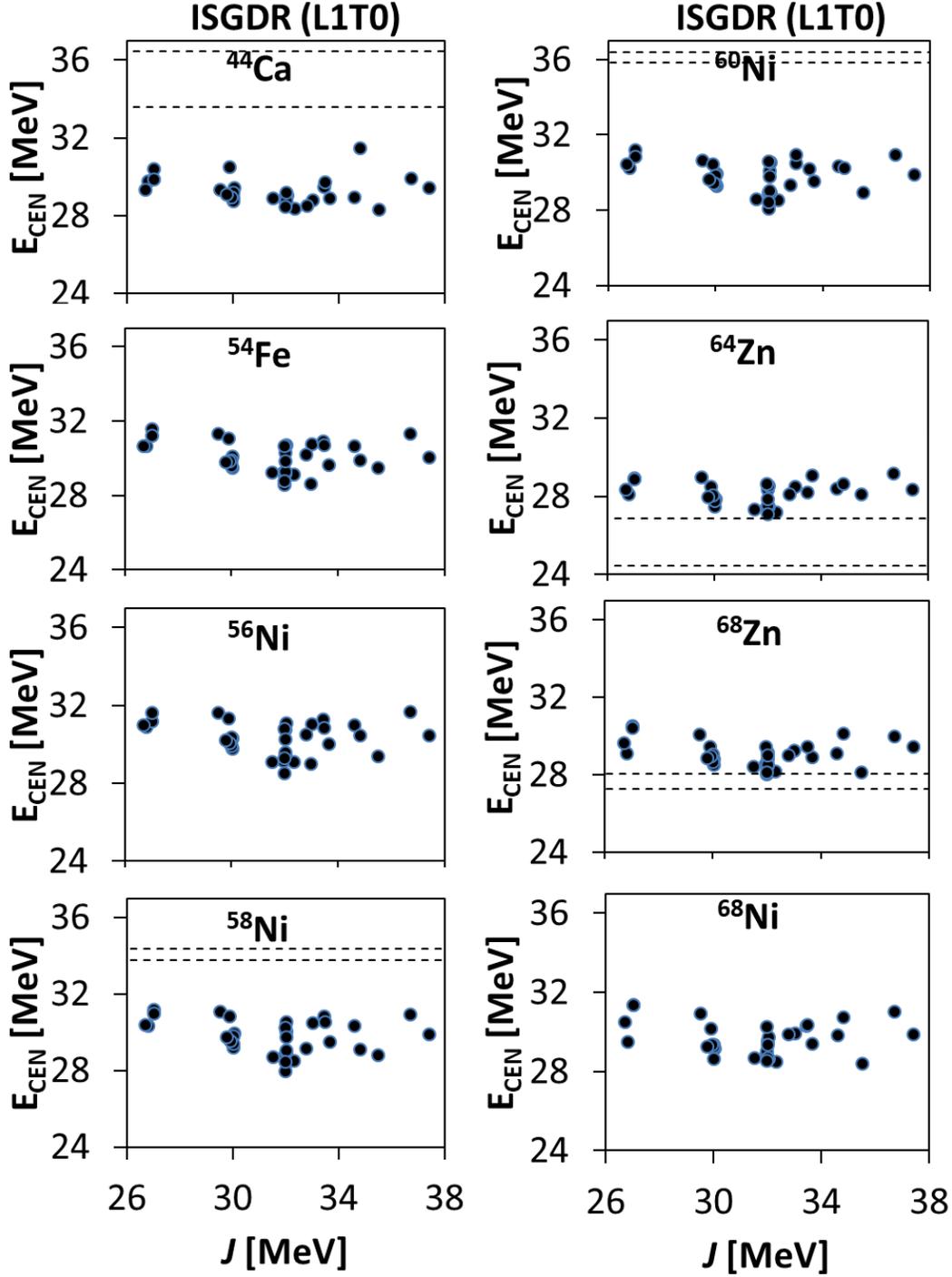

FIG. 8. Similar to FIG. 1, for the ISGDR as a function of the symmetry energy $J$ at saturation density for different nuclei. We find no correlation between the calculated values of $E_{CEN}$ and $J$ (Pearson linear correlation coefficient C ~ -0.17).

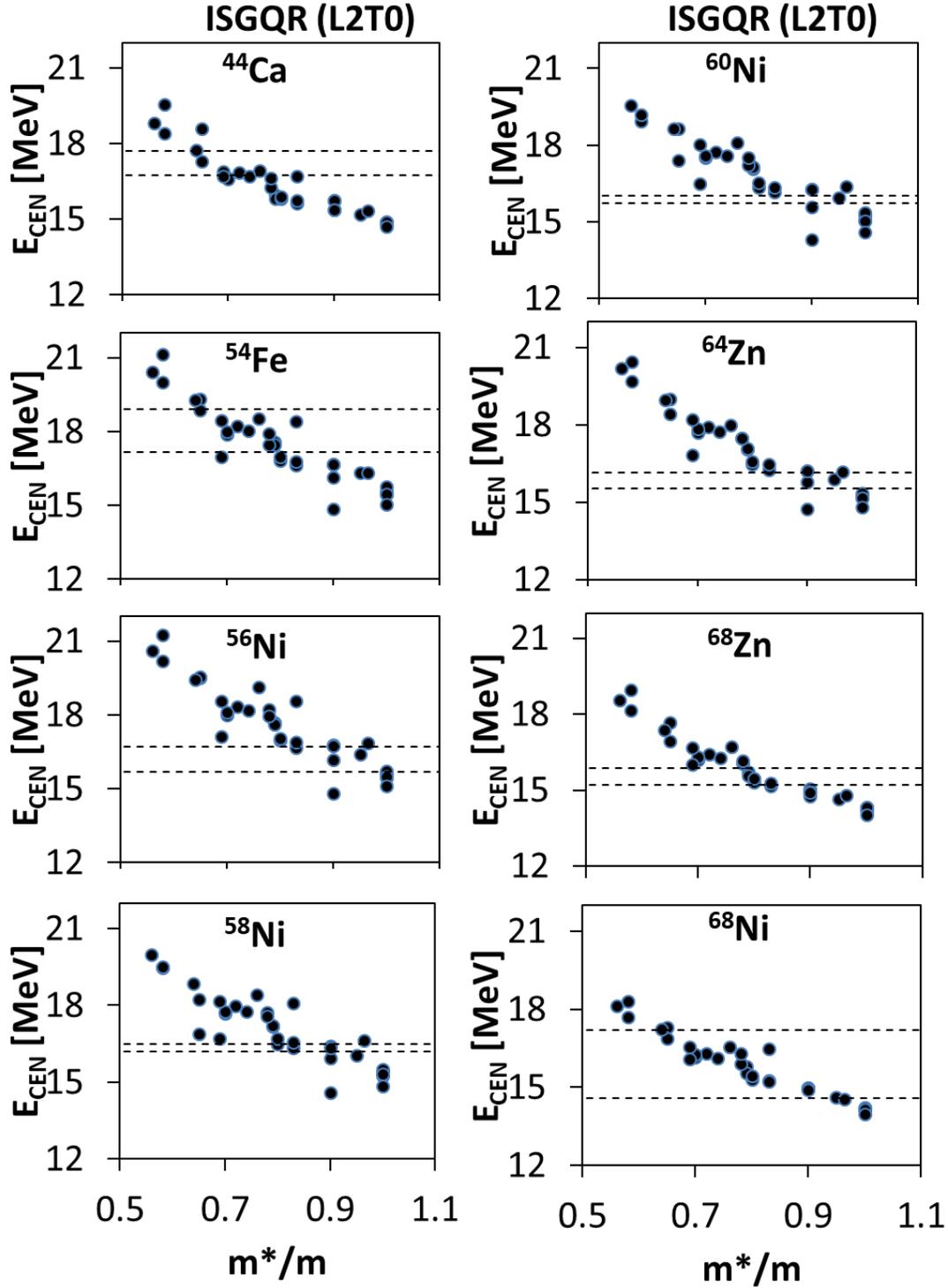

FIG. 9. Similar to FIG. 1, for the ISGQR as a function of the effective mass $m^*/m$. We find strong correlation between the calculated values of $m^*/m$ and $E_{CEN}$ with a Pearson linear correlation coefficient $C \sim -0.93$ in all cases.

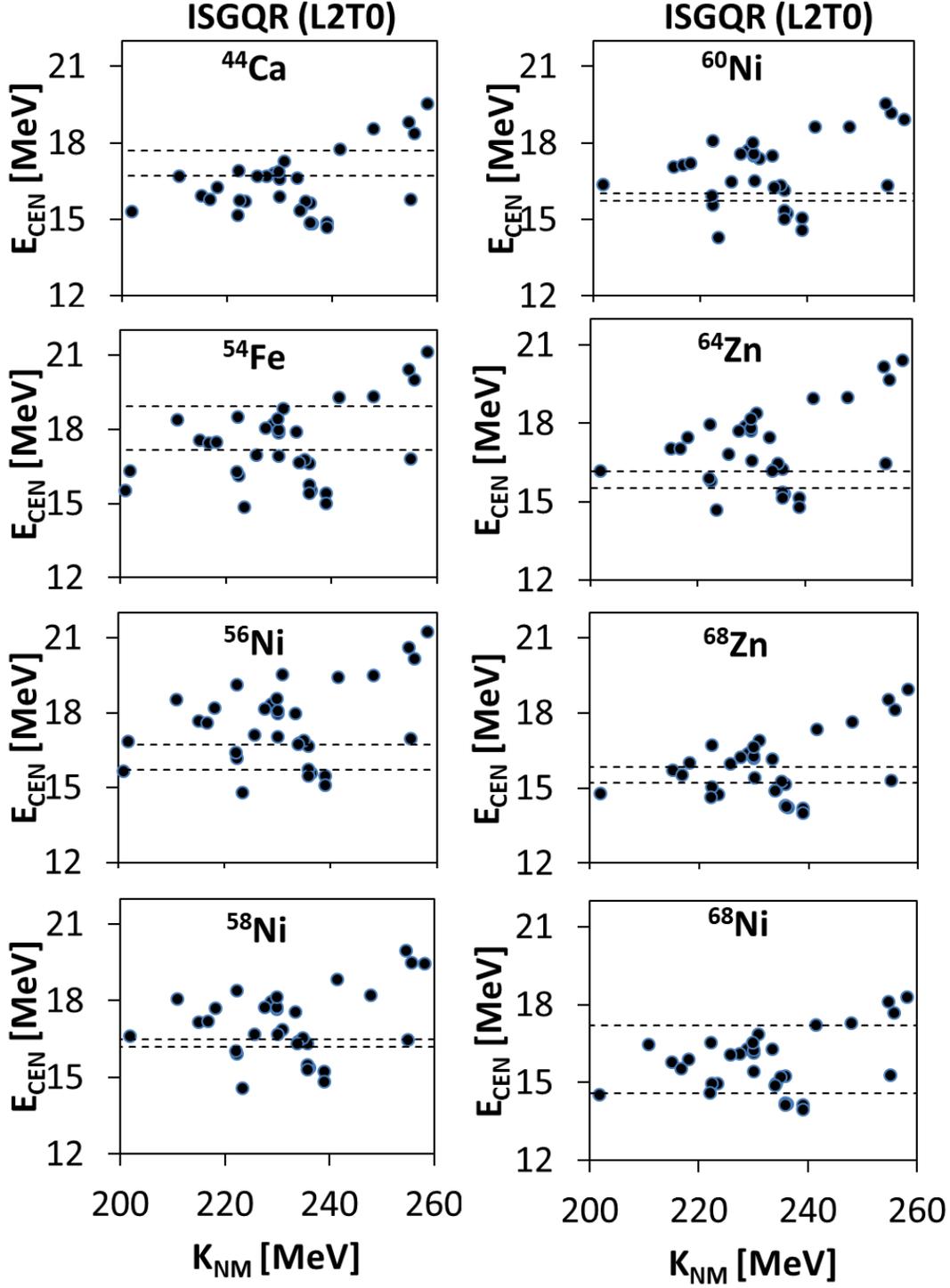

FIG. 10. Similar to FIG. 1, for the ISGQR as a function of the incompressibility coefficient $K_{NM}$. We find weak correlation between the calculated values of $K_{NM}$ and $E_{CEN}$ with a Pearson linear correlation coefficient close to C ~ 0.40 for all isotopes, mostly due to the correlation between $K_{NM}$ and $m^*/m$ shown in Table V of Ref. [11].

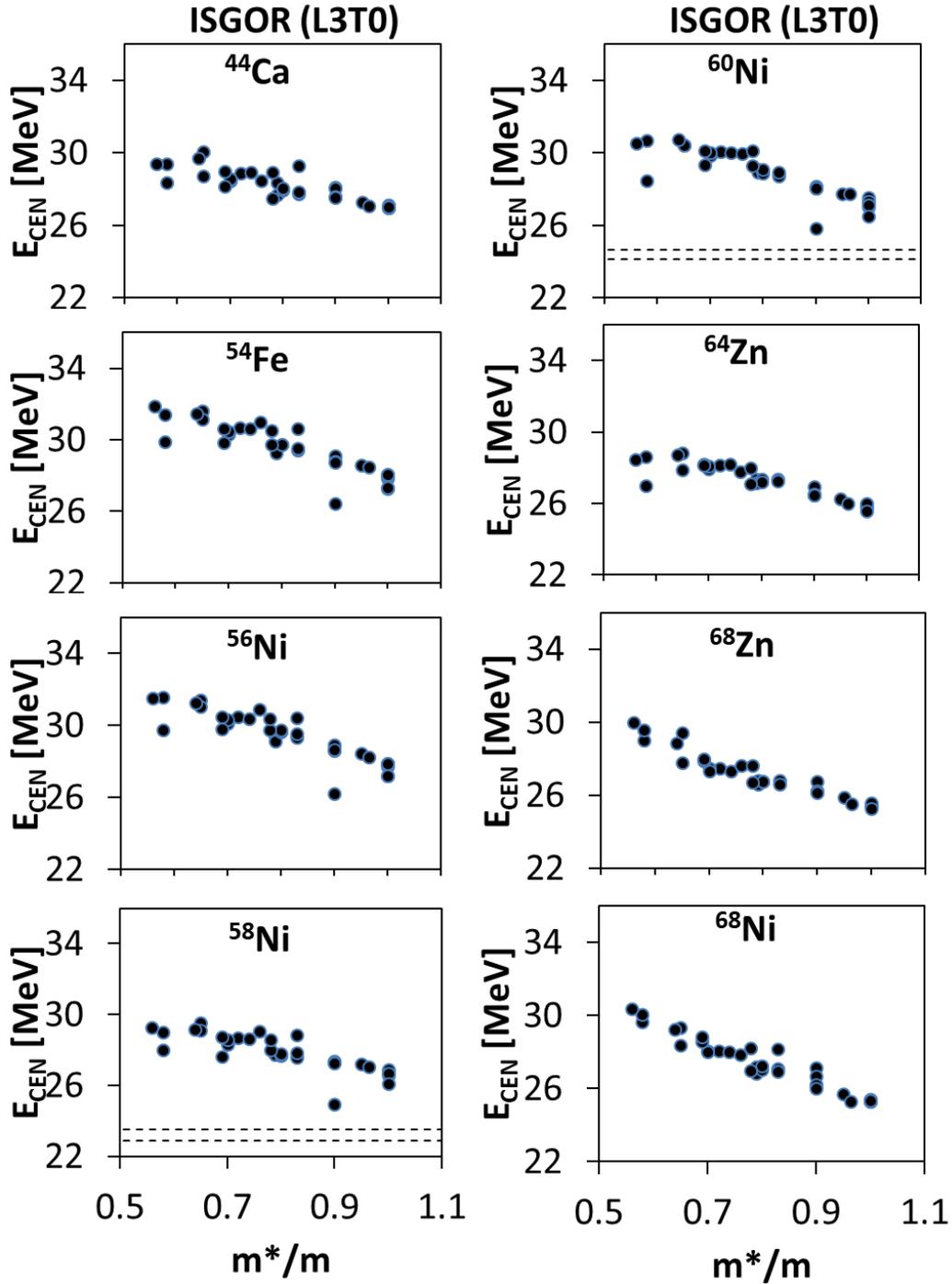

FIG. 11. Similar to FIG. 1, for the ISGOR plotted against the effective mass m*/m. We find strong correlation between the calculated values of m*/m and $E_{CEN}$ with a Pearson linear correlation coefficient C greater in magnitude than -0.89 in all cases.

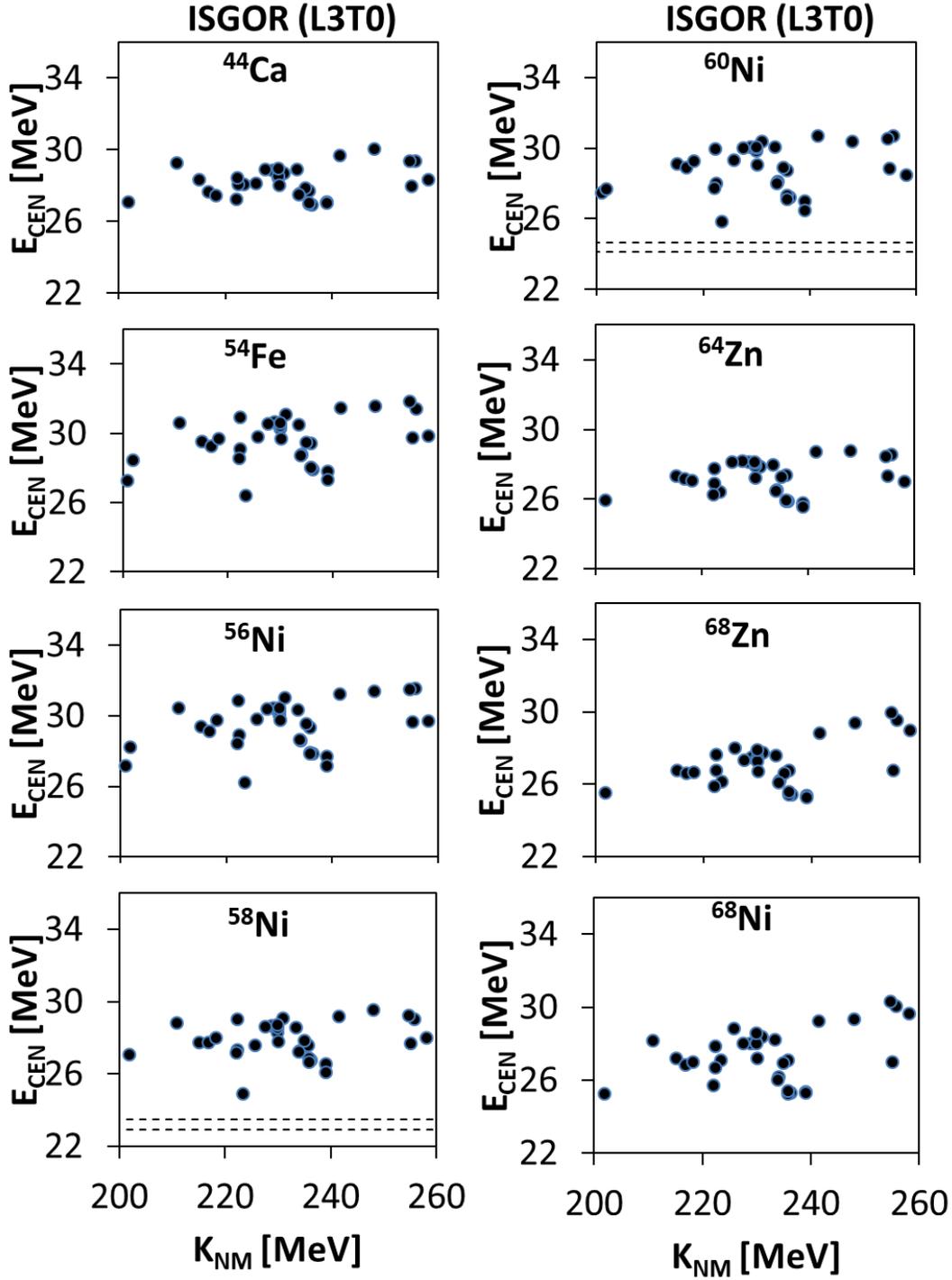

FIG. 12. Similar to FIG. 1, for the ISGOR plotted against the incompressibility coefficient $K_{NM}$. We find no correlation between the calculated values of $K_{NM}$ and $E_{CEN}$ with a Pearson linear correlation coefficient roughly C ~ 0.32 in all cases.

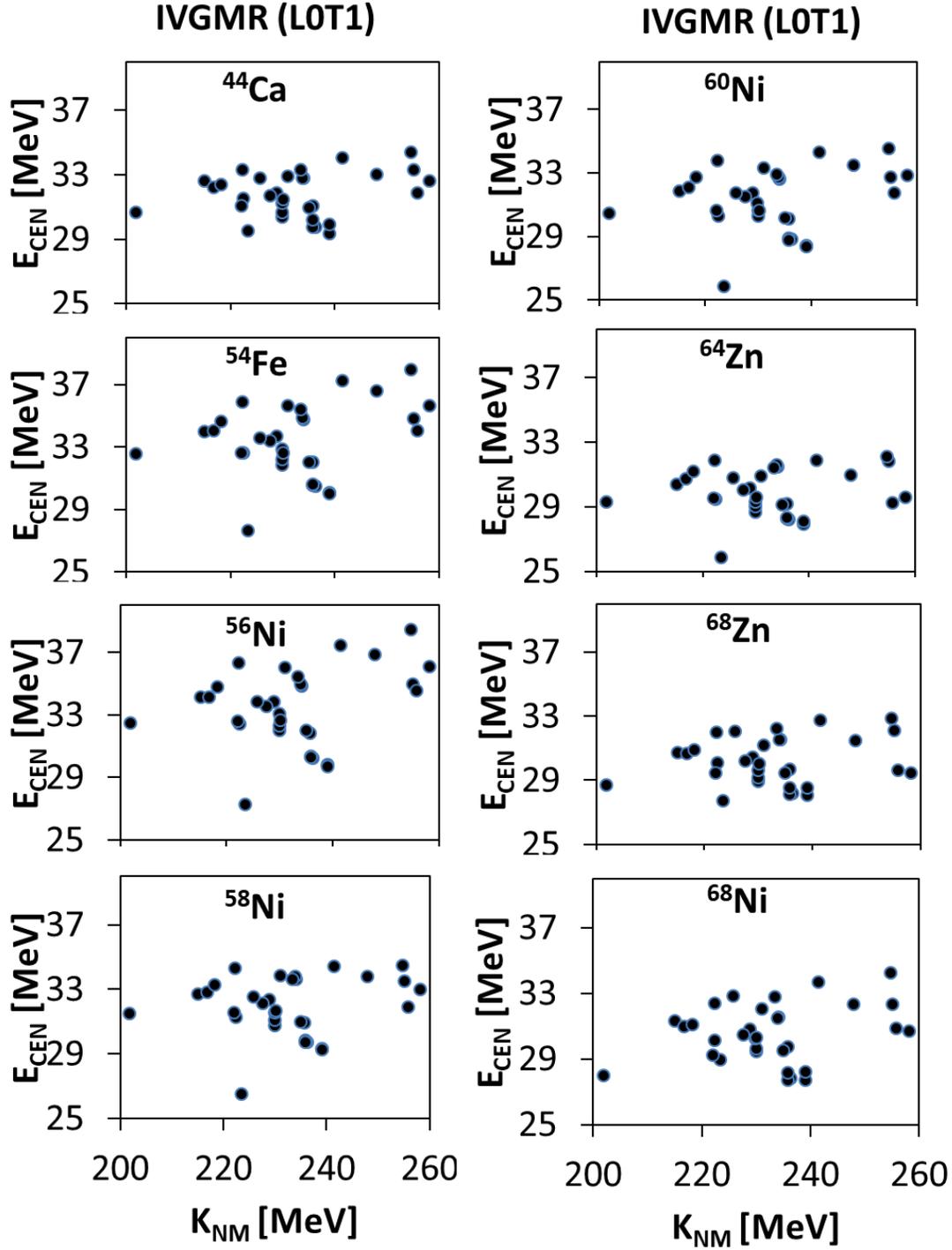

FIG. 13. Similar to FIG. 1, for the IVGMR plotted against the incompressibility coefficient of nuclear matter $K_{NM}$. We do not find any correlation between the calculated values of $E_{CEN}$ and $K_{NM}$ with a Pearson linear correlation coefficient C ~ 0.22.

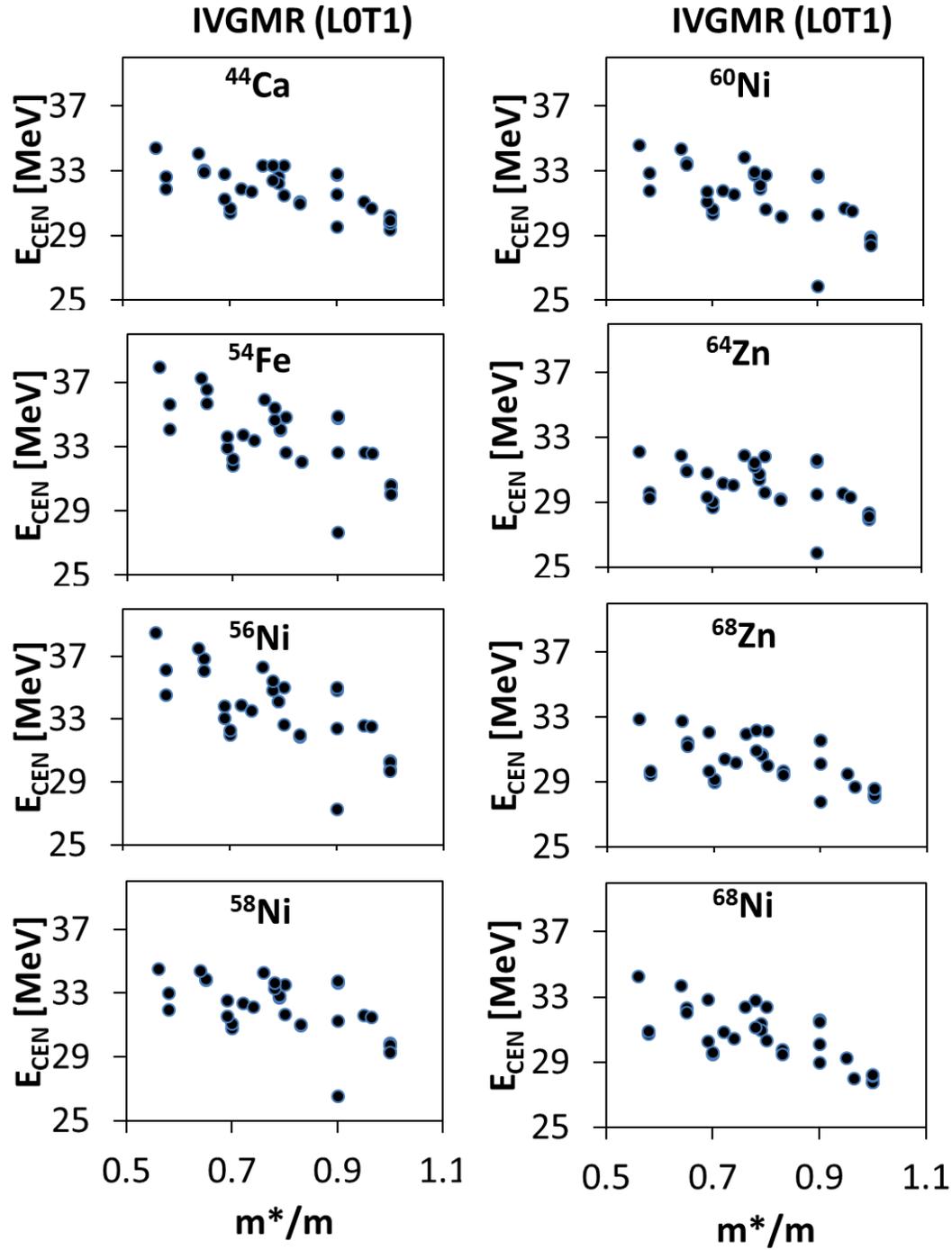

FIG. 14. Similar to FIG. 1, for the IVGMR plotted against the effective mass m*/m. We find medium correlation between the calculated values of $E_{CEN}$ and m*/m with a Pearson linear correlation coefficient C ~ -0.64 for all the isotopes considered here.

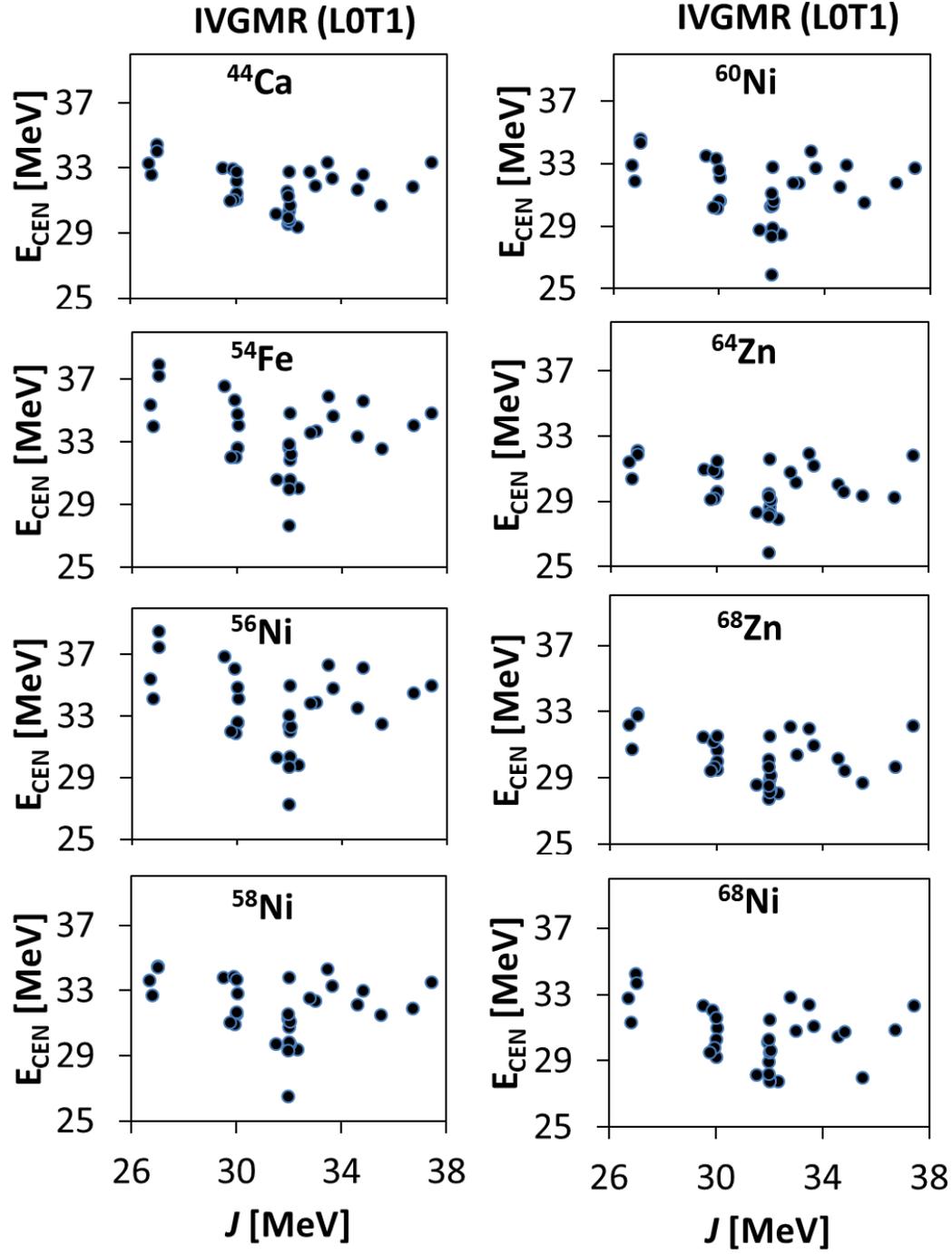

FIG. 15. Similar to FIG. 1, for the IVGMR as a function of the symmetry coefficient $J$ at saturation density. We do not find any correlation between the value of $J$ and $E_{CEN}$ with a Pearson linear correlation coefficient C ~ -0.24.

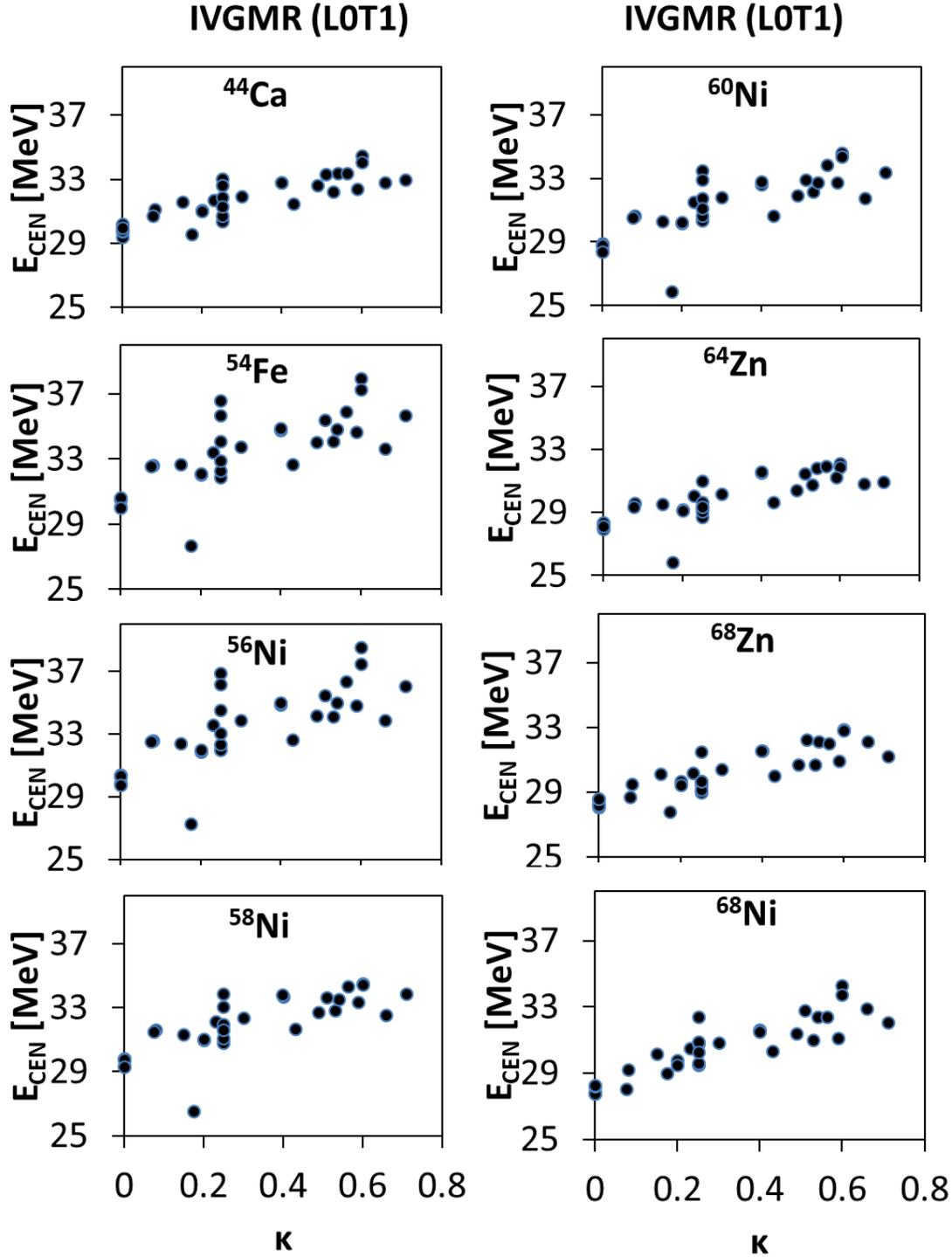

FIG. 16. Similar to FIG. 1, for the IVGMR plotted against the enhancement coefficient, κ, of the energy weighted sum rule (EWSR) of the isovector giant dipole resonance (IVGDR). We find strong correlation between the calculated value of κ and $E_{CEN}$ with a Pearson linear correlation coefficient C ~ 0.80 for all isotopes considered.

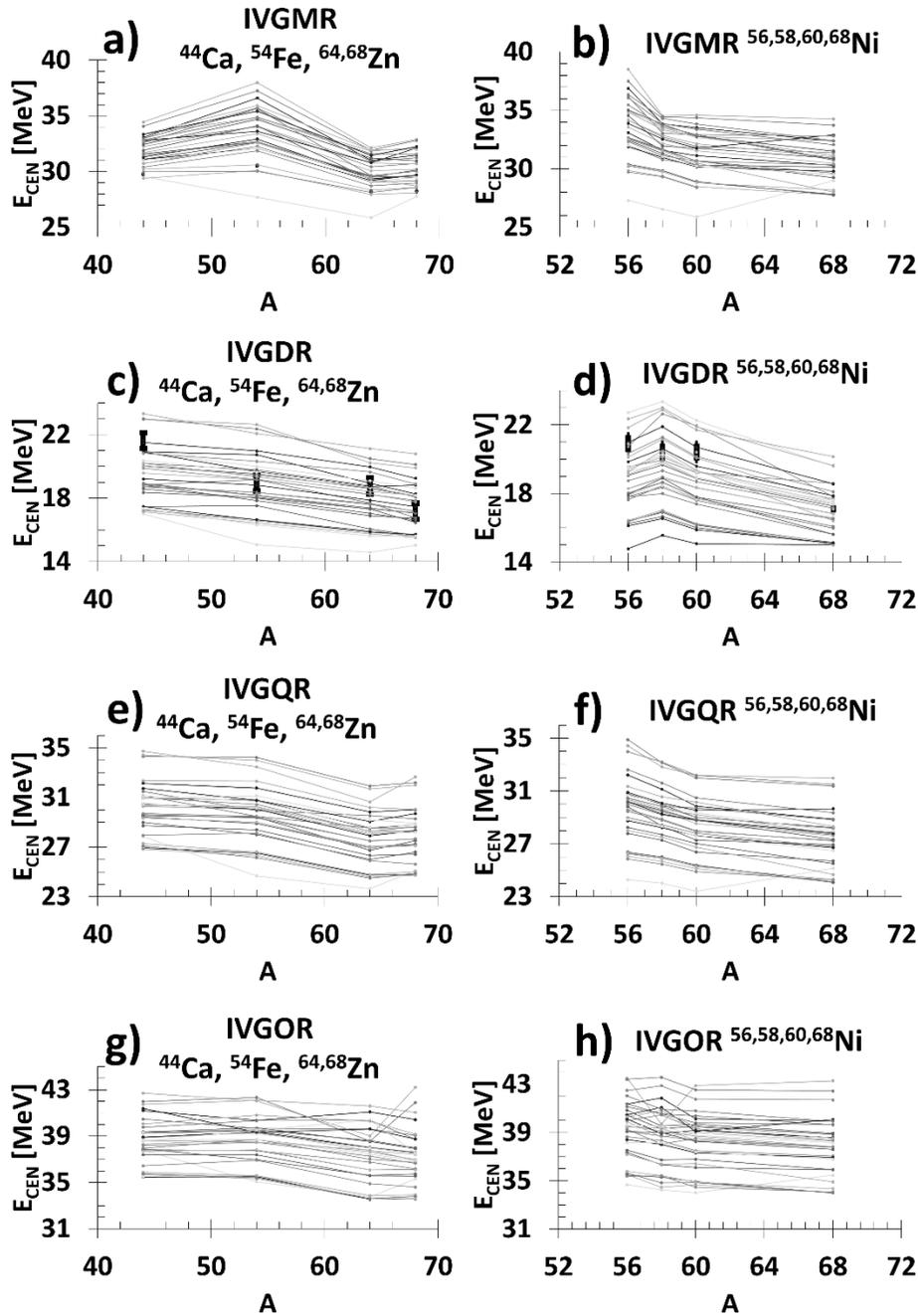

FIG. 17. The centroid energies [MeV] of the isovector giant resonances of multipolarities L=0-3 for $^{44}$Ca, $^{54}$Fe, and $^{64,68}$Zn (left figures) and for $^{56,58,60,68}$Ni (right figures), are plotted against the mass A of each isotope. The experimental error bar (where available) are shown by the solid vertical lines, while the theoretical values of $E_{CEN}$ are shown as dots connected by lines (to guide the eye).

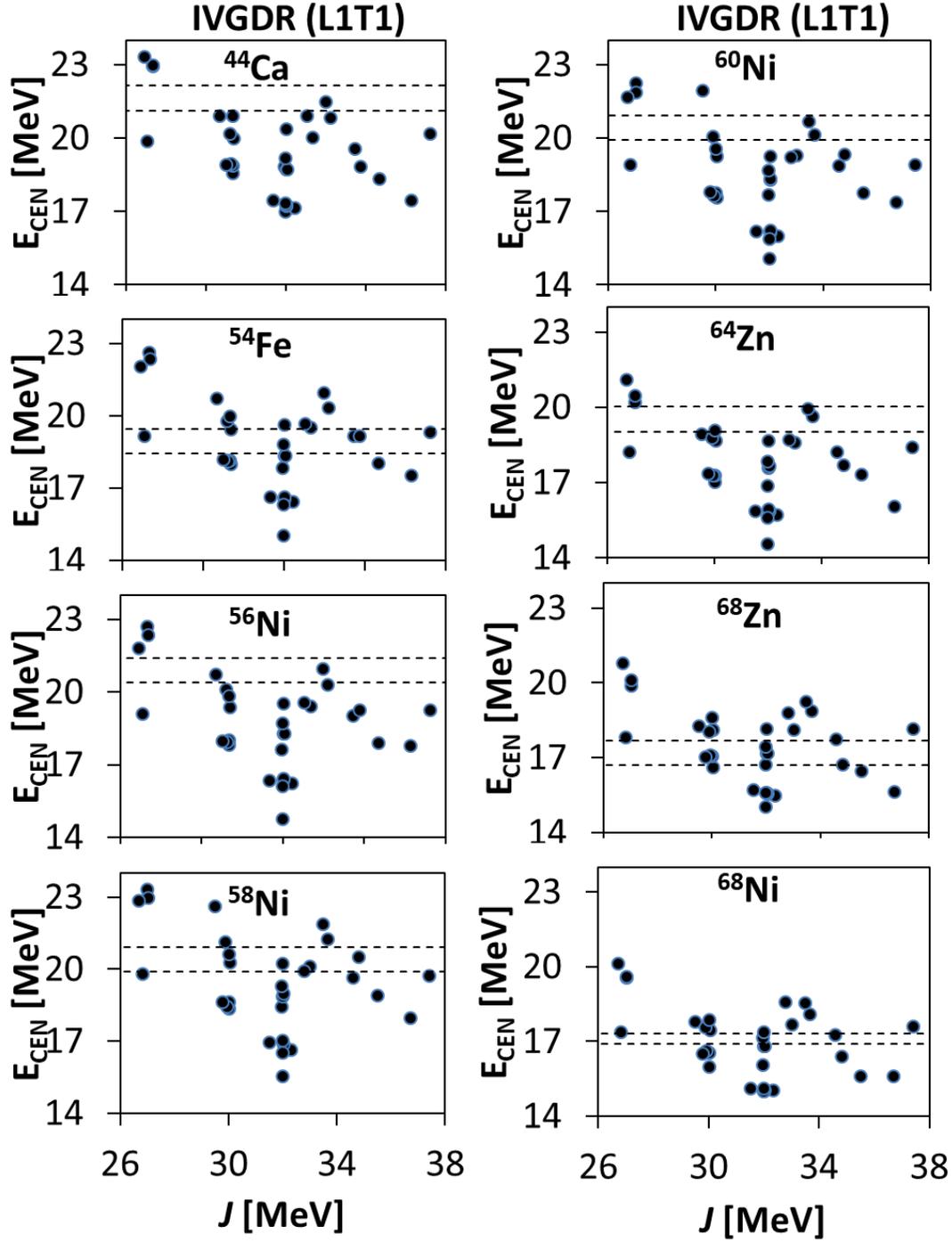

FIG. 18. Similar to FIG. 1, for the IVGDR as a function of the symmetry energy $J$ at saturation density. We find weak correlation between the value of $J$ and the value of $E_{CEN}$ with a Pearson linear correlation coefficient C ~ -0.39.

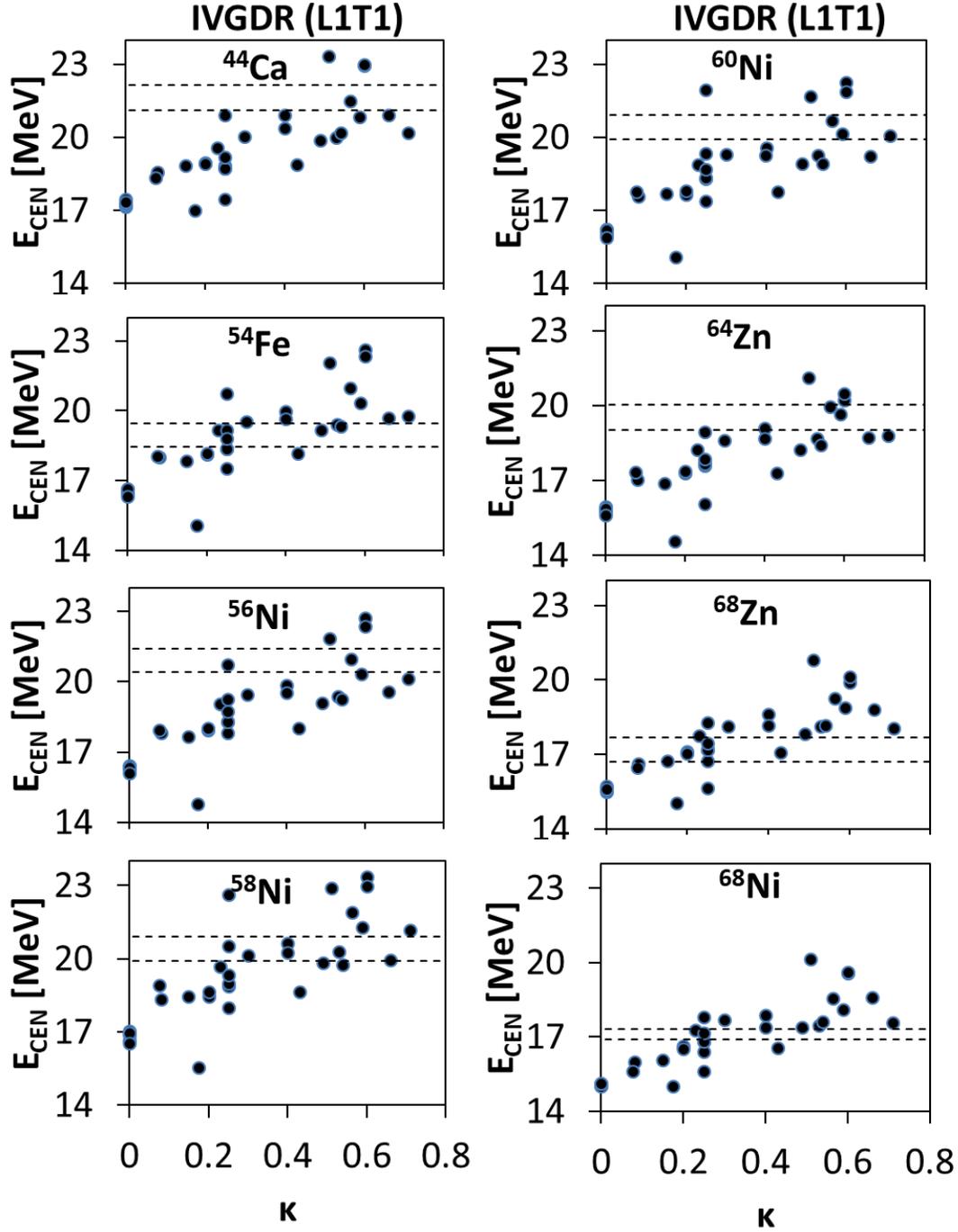

FIG. 19. Similar to FIG. 1, for the IVGDR for different nuclei plotted against the energy weighted sum rule enhancement coefficient κ of the IVGDR. We find strong correlation between the calculated values of κ and $E_{CEN}$ with a Pearson linear correlation coefficient C ~ 0.80 for all isotopes considered.

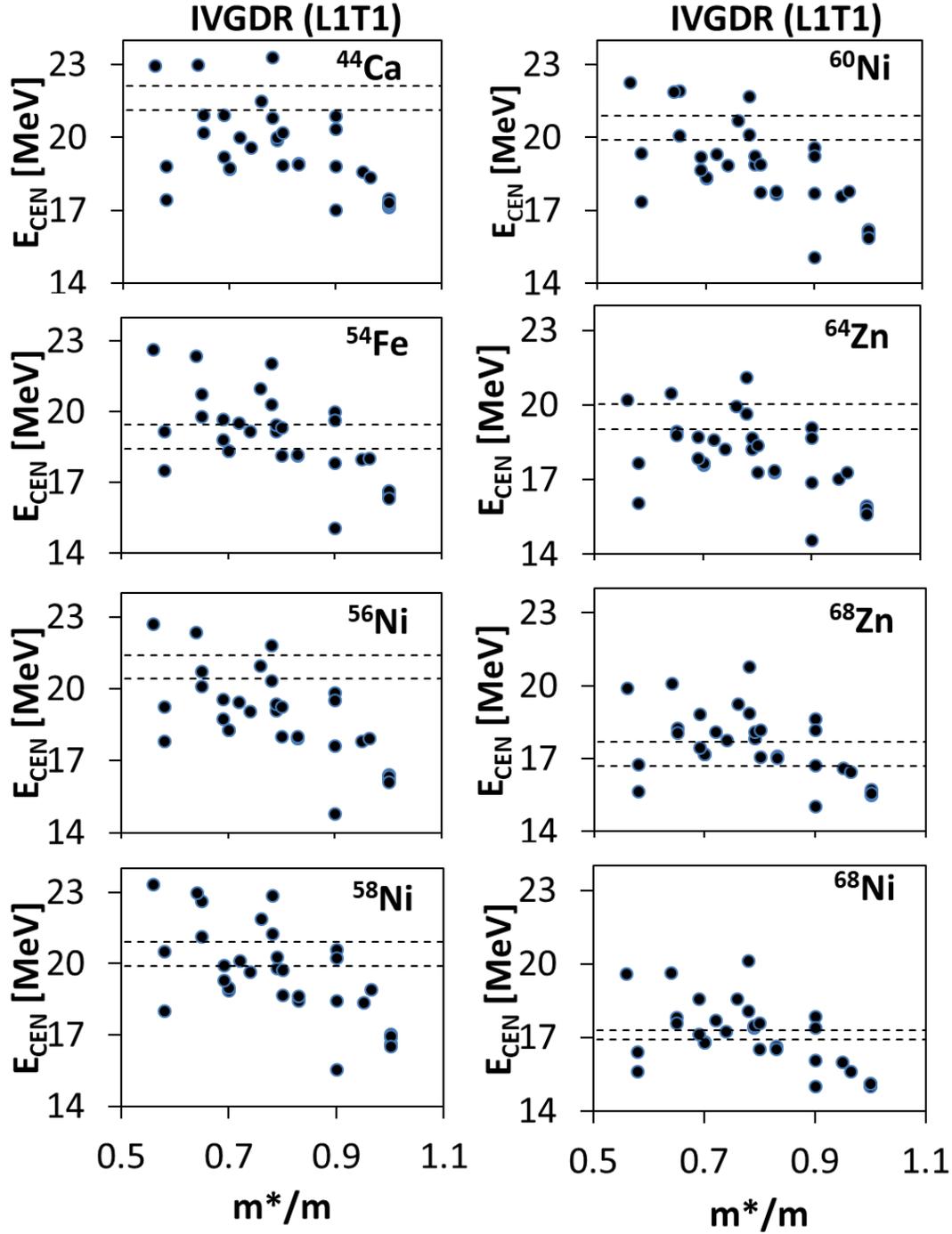

FIG. 20. Similar to FIG. 1, for the IVGDR plotted against the effective mass m*/m. We find medium correlation between the calculated values of $E_{CEN}$ and m*/m with a Pearson linear correlation coefficient close to C ~ -0.62 for all the isotopes considered here.

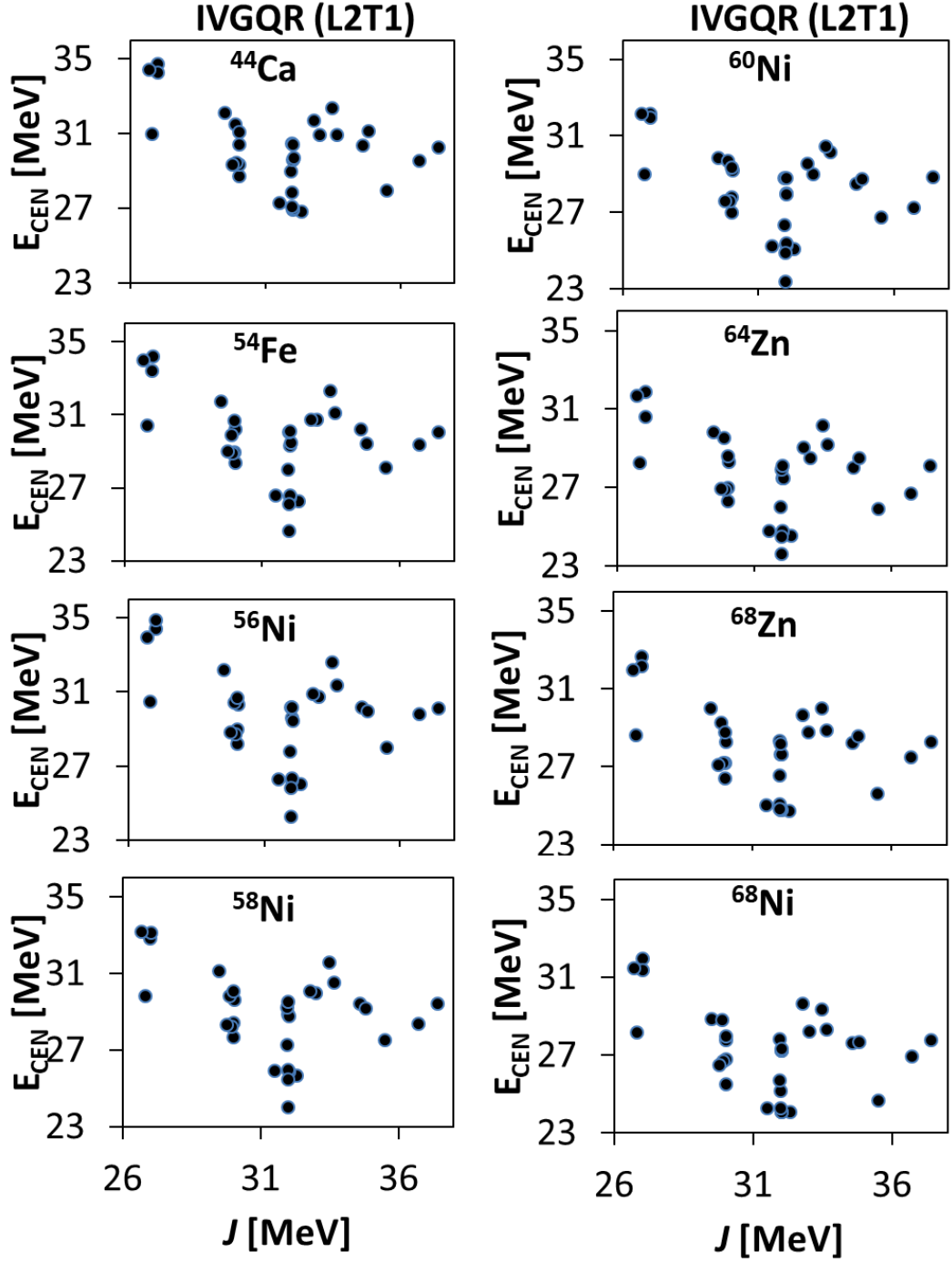

FIG. 21. Similar to FIG. 1, for the IVGQR as a function of the symmetry energy $J$ at saturation density. We find a weak correlation between the value of $J$ and the value of $E_{CEN}$, with a Pearson linear correlation coefficient of C ~ -0.38.

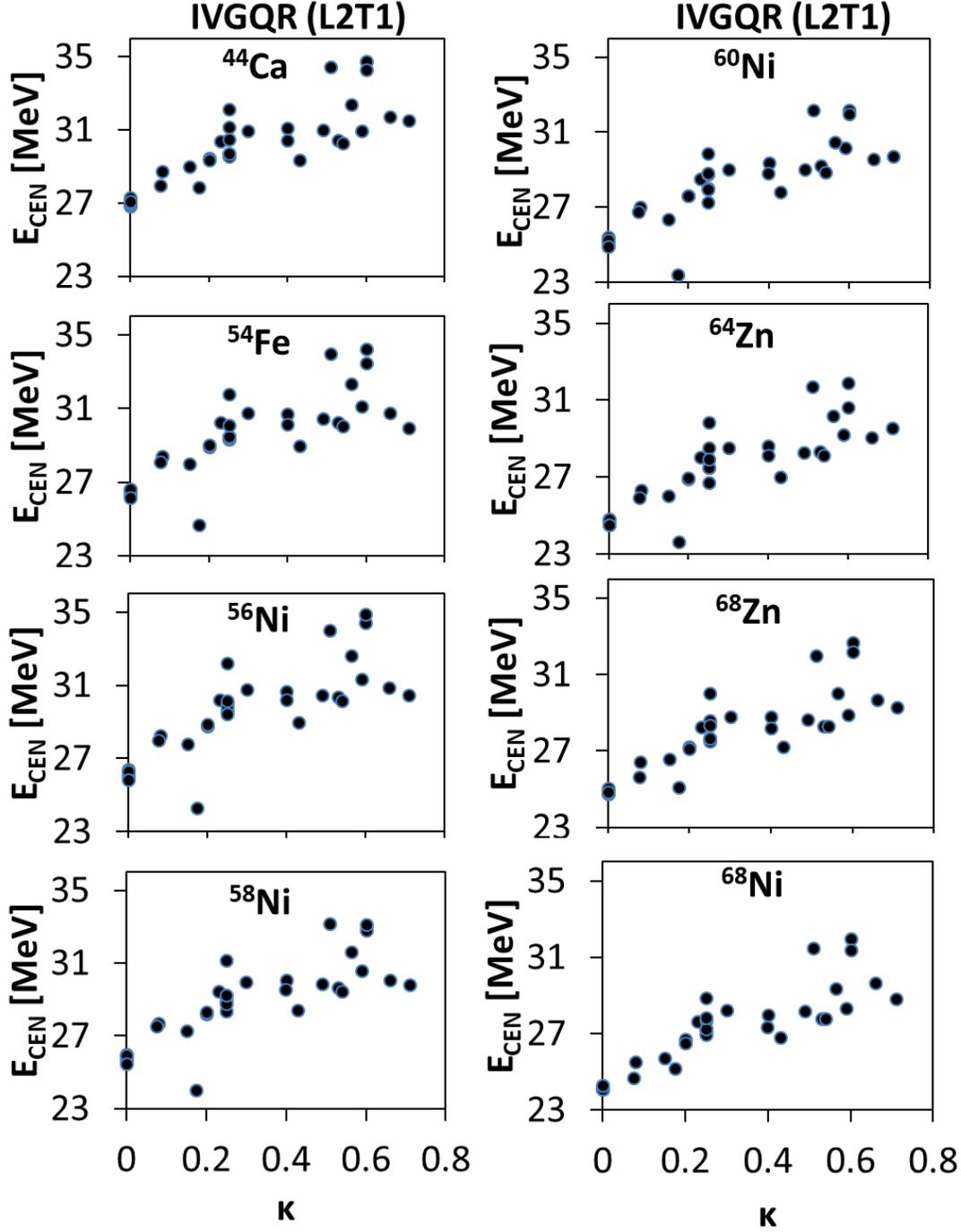

FIG. 22. Similar to FIG. 1, for the IVGQR plotted against the enhancement coefficient κ of the EWSR of the IVGDR. We find a strong correlation between the calculated values of κ and $E_{CEN}$ with a Pearson linear correlation coefficient C ~ 0.81 for all isotopes considered.

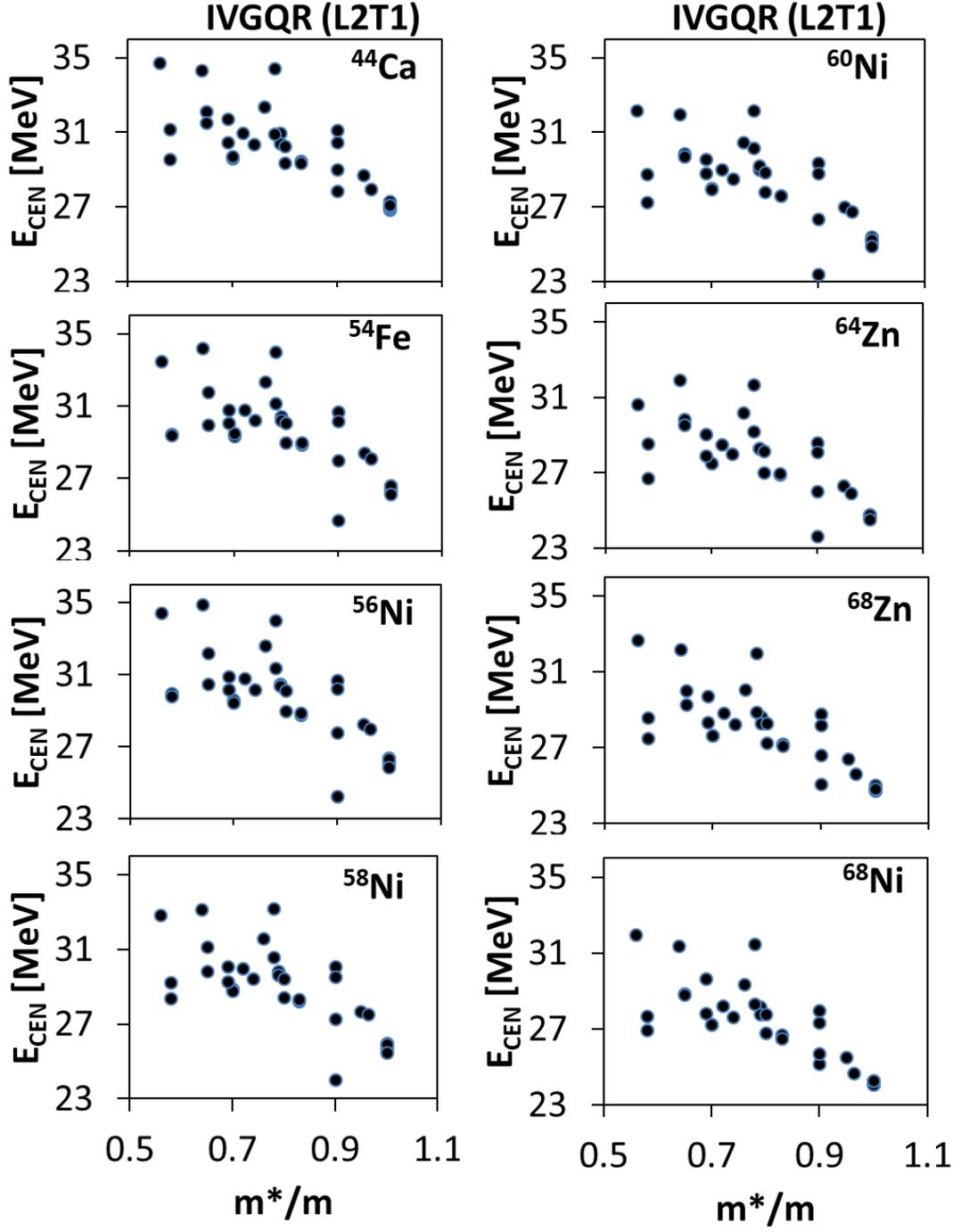

FIG. 23. Similar to FIG. 1, for the IVGQR of different nuclei plotted against the effective mass m*/m. We find medium correlation between the calculated values of $E_{CEN}$ and m*/m, with a Pearson linear correlation coefficient close to C ~ -0.73 for all the isotopes considered here.

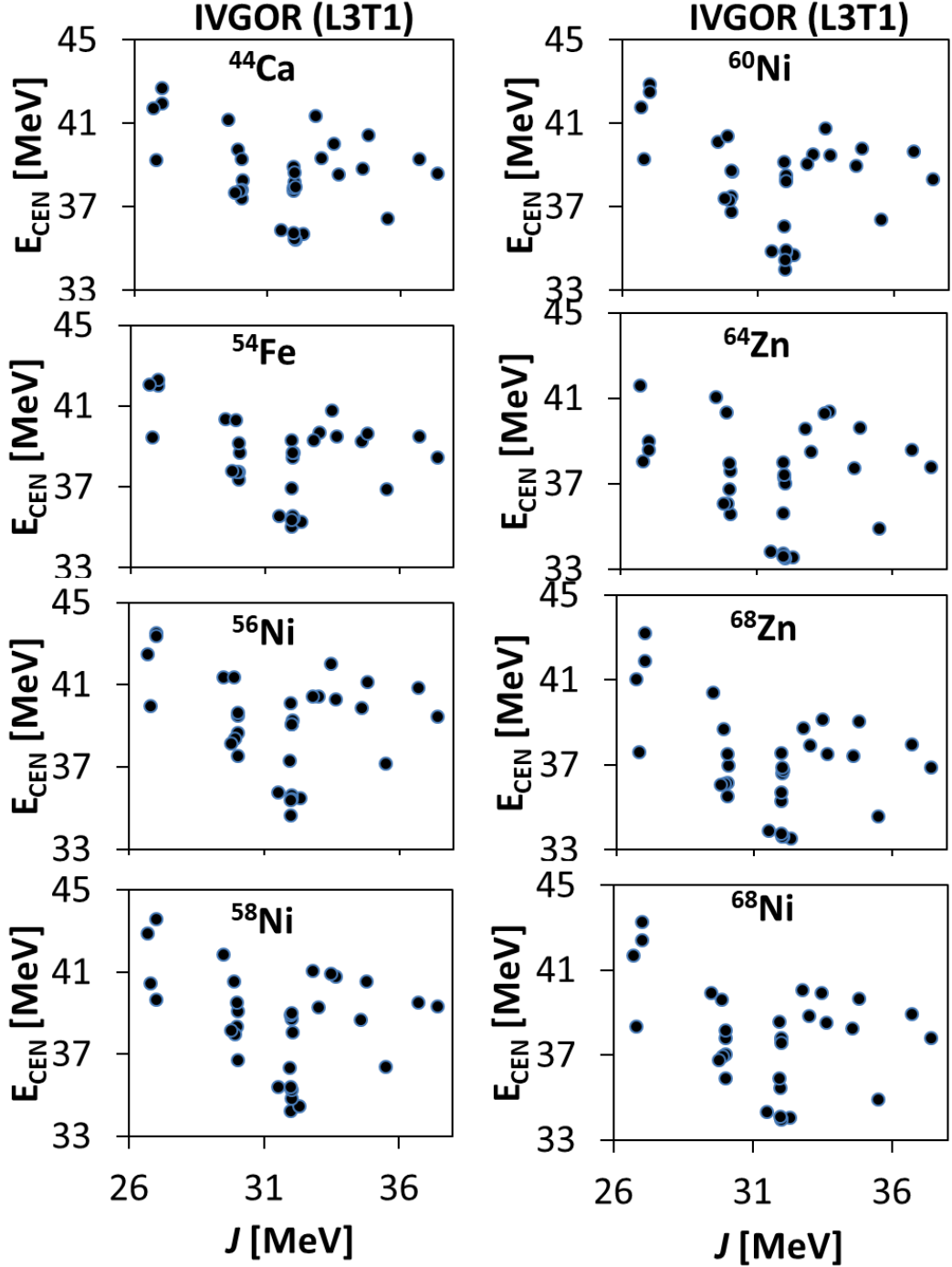

FIG. 24. Similar to FIG. 1, for the IVGOR as a function of the symmetry energy $J$ at saturation density. We do not find any correlation between the value of $J$ and the value of $E_{CEN}$ with a Pearson linear correlation coefficient C ~ -0.29.

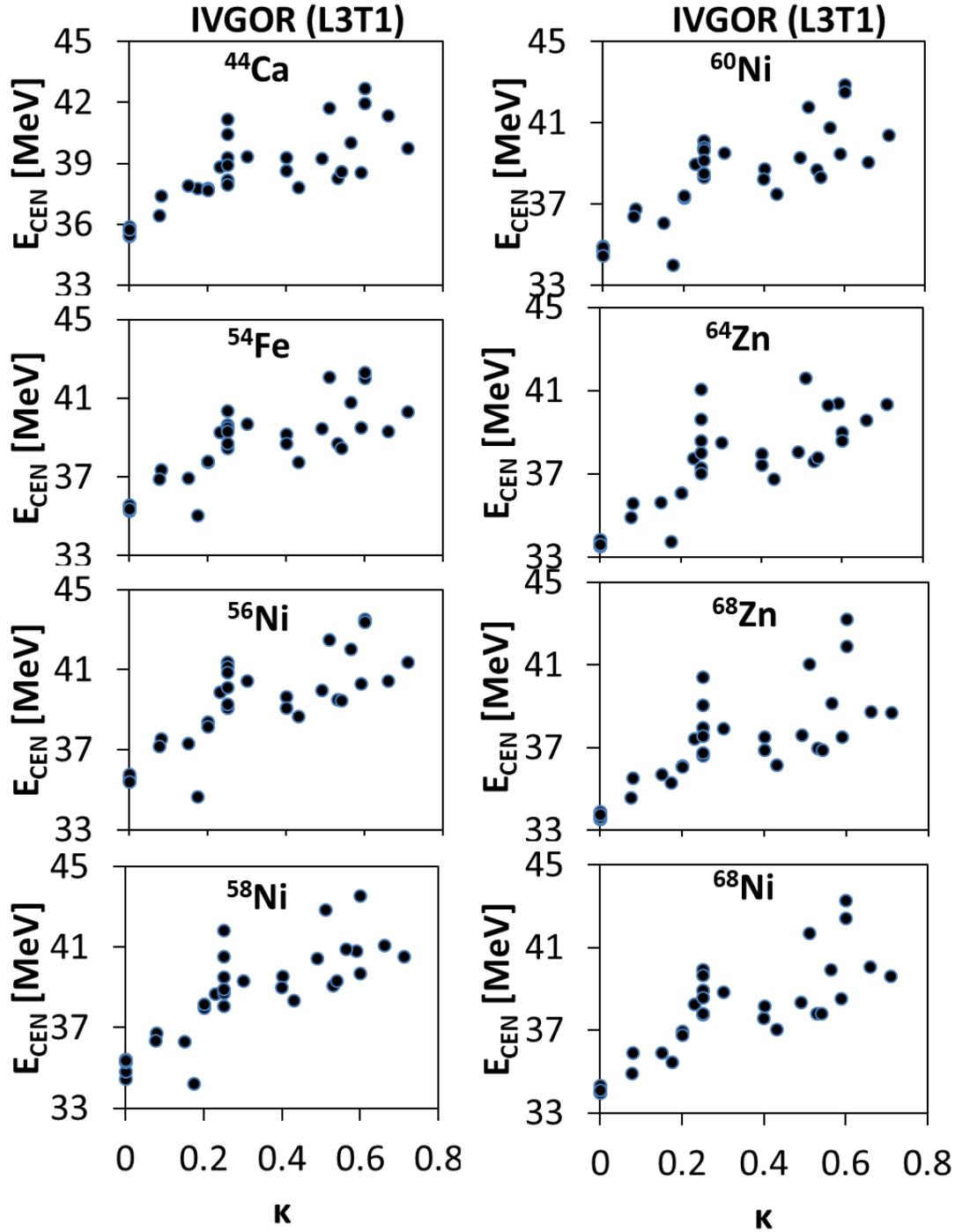

FIG. 25. Similar to FIG. 1, for the IVGOR plotted against the enhancement coefficient κ of the EWSR of the IVGDR. We find a medium correlation between the values of κ and $E_{CEN}$ with a Pearson linear correlation coefficient C ~ 0.79 for all isotopes considered.

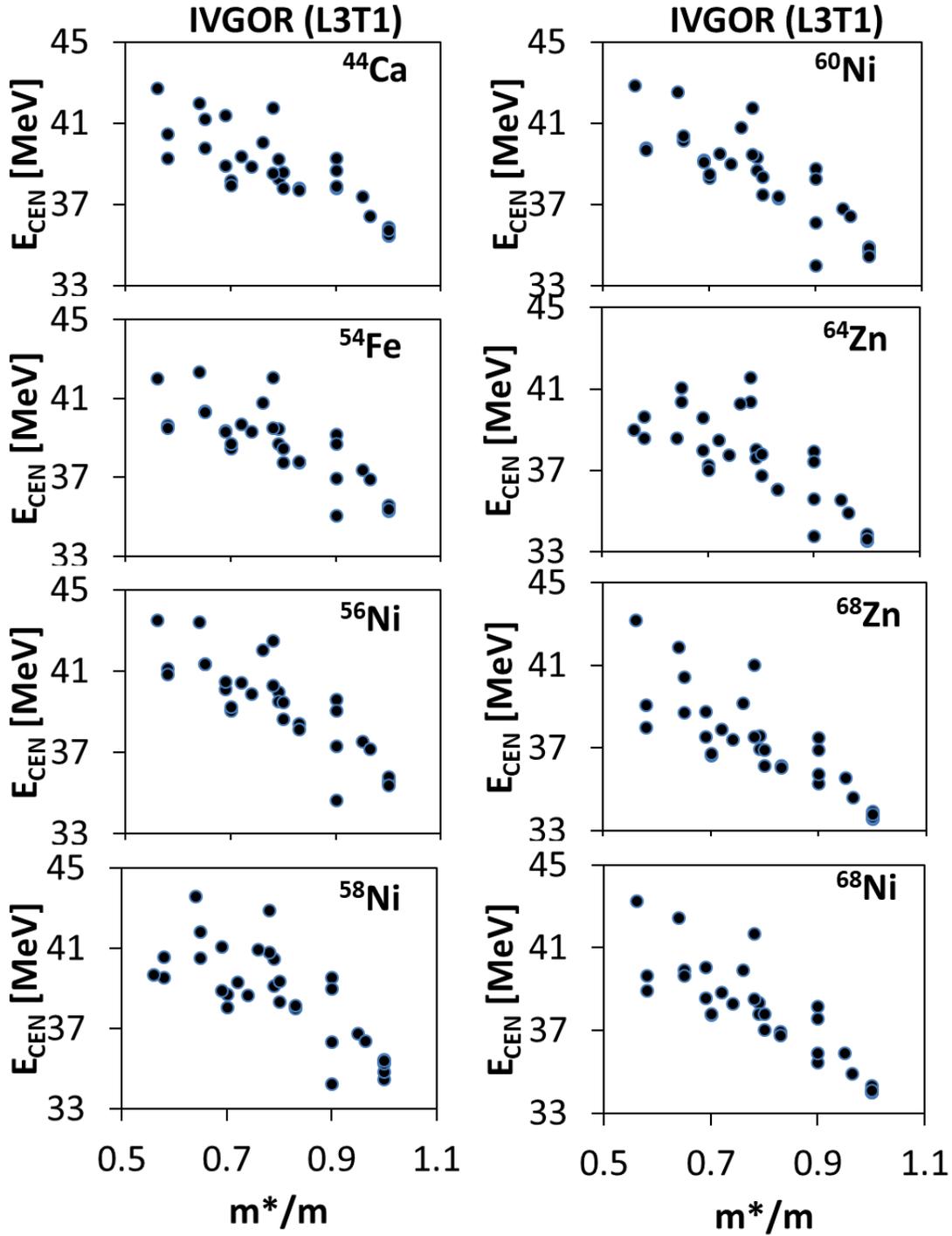

FIG. 26. Similar to FIG. 1, for the IVGOR plotted against the effective mass m*/m. We find strong correlation between the calculated values of m*/m and $E_{CEN}$ with a Pearson linear correlation coefficient C ~ -0.82 for all isotopes considered.